\documentclass[12pt]{article}
\usepackage[centertags]{amsmath}
\usepackage{amsfonts}
\usepackage{amssymb}
\usepackage{amsthm}
\usepackage{newlfont}
\usepackage{epsfig}
\usepackage{amscd}

\newcommand{\NN}{{\mathbb N}}

\newcommand{\CC}{{\mathbb C}}
\newcommand{\beq}{\begin{equation}}
\newcommand{\eeq}{\end{equation}}
\newcommand{\ba}{\begin{array}}
\newcommand{\ea}{\end{array}}
\newcommand{\bea}{\begin{eqnarray}}
\newcommand{\eea}{\end{eqnarray}}

\begin{document}

\begin{center}
{\large \sc \bf On the remarkable relations among PDEs integrable by \\
the inverse spectral transform method, by
the method of characteristics and by the Hopf-Cole transformation}

\vskip 15pt

{\large A. I. Zenchuk$^{1,\S}$ and P. M. Santini$^{2,\S}$ }

\vskip 8pt

{\it $^1$ Center of Nonlinear Studies of the Landau Institute for
Theoretical Physics \\
(International Institute of Nonlinear Science)\\  Kosygina 2,
Moscow, Russia 119334}

\smallskip

{\it
$^2$ Dipartimento di Fisica, Universit\`a di Roma "La Sapienza" and\\
Istituto Nazionale di Fisica Nucleare, Sezione di Roma 1\\
Piazz.le Aldo Moro 2, I-00185 Roma, Italy}

\vskip 5pt

$^{\S}$e-mail:  {\tt zenchuk@itp.ac.ru , paolo.santini@roma1.infn.it}

\vskip 5pt

{\today}

\end{center}

\begin{abstract}
We establish deep and remarkable connections among partial differential equations (PDEs) integrable by 
different methods: the inverse spectral transform method, the method of characteristics 
and the Hopf-Cole transformation. More concretely, 
1) we show that the integrability properties (Lax pair, infinitely-many commuting 
symmetries, large classes of analytic solutions) of 
(2+1)-dimensional PDEs integrable by the 
Inverse Scattering Transform method ($S$-integrable) can be generated by 
the integrability properties of the (1+1)-dimensional matrix B\"urgers hierarchy, integrable by the 
matrix Hopf-Cole transformation ($C$-integrable). 
2) We show that the integrability properties i) of $S$-integrable PDEs in (1+1)-dimensions, ii) of
the multidimensional   
generalizations of the $GL(M,\CC)$ self-dual Yang Mills equations, and iii) of the multidimensional Calogero 
equations can be generated by the integrability properties of a recently introduced multidimensional matrix equation solvable 
by the method of characteristics. To establish the above links, we consider a 
block Frobenius matrix reduction of the relevant matrix fields, leading to integrable chains of matrix equations for the 
blocks of such a Frobenius matrix, followed by a systematic elimination procedure of some of these blocks. 
The construction of large classes 
of solutions of the soliton equations from solutions of the matrix B\"urgers hierarchy turns 
out to be intimately related to the construction of solutions in Sato theory. 3) We finally show 
that suitable generalizations of the block Frobenius matrix reduction of the matrix B\"urgers hierarchy 
generates PDEs exhibiting integrability properties in common with both $S$- and $C$- integrable equations.

\end{abstract}

\section{Introduction}
Integrable nonlinear partial differential equations (PDEs) can be grouped into different classes,  
depending on their method of solution. We distinguish the following three basic classes.  

\begin{enumerate}
\item
Equations solvable by the method of characteristics \cite{Whitham}, hereafter called, 
for the sake of brevity, $Ch$-integrable, like the following 
matrix PDE in arbitrary dimensions \cite{SZ}:
\begin{eqnarray}\label{w}
w_{t} + \sum_{i=1}^N w_{x_i}\rho^{(i)}(w) + [B,w]\sigma(w)=0,
\end{eqnarray}
where $w$ is a square matrix and $\rho^{(i)}(\cdot ),\sigma(\cdot )$ are scalar functions 
representable as positive power series, or like the  
vector equations solvable by the generalized hodograph method \cite{ts1, dn, ts2, F, K1, K2}.

\item
Equations integrable by a simple change of variables, often called $C$-integrable \cite{Calogero}, 
like the matrix B\"urgers equation \cite{BLR}
\begin{eqnarray}\label{Burgers0}
w_{t} - B w_{xx} -2Bw_x w +[w,B] (w_x+w^2) =0,
\end{eqnarray}
where $B$ is any constant square matrix, linearizable by the matrix version of the Hopf-Cole transformation 
$\Psi_x=w\Psi$ 
\cite{Hopf}.

\item
Equations integrable by less elementary methods of spectral nature, the inverse spectral 
transform (IST) \cite{GGKM, ZMNP, CD, AC, Konop} and the dressing method \cite{ZS1, ZS2, ZM, BM, Konop}, 
often called $S$-integrable \cite{Calogero} or soliton equations. Within this class of equations, 
we distinguish four 
different subclasses, depending on the nature of the associated spectral theory. 
\begin{enumerate}
\item
Soliton equations in (1+1)-dimensions like, for instance, the Korteweg-de Vries 
(KdV) \cite{KdV,GGKM} and the Nonlinear Schr\"odinger (NLS) \cite{ZS_NLS} equations, whose inverse 
problems are local Riemann-Hilbert (RH) problems \cite{ZMNP,AC}.
\item
Their (2+1)-dimensional generalizations, like the Kadomtsev-Petviashvily (KP) \cite{KP} 
and Davey-Stewartson (DS) \cite{DS} equations, whose inverse problems are nonlocal RH 
\cite{Manakov_KP, FA} or $\bar\partial$ - problems \cite{ABF}.
\item
The self-dual Yang-Mills (SDYM) equation \cite{Ward,BZ} and its generalizations in arbitrary dimensions. 
\item
Multidimensional PDEs associated with one-parameter families of commuting vector fields, whose novel IST,  
recently constructed in \cite{MS1, MS2}, is characterized by nonlinear RH \cite{MS1, MS2} or 
$\bar\partial$ \cite{K-MA-R} problems. Distinguished 
examples are the dispersion-less KP equation, the heavenly equation of Plebanski \cite{heavenly} and the 
following integrable system of PDEs in $N+4$ dimensions \cite{MS1}:
\begin{eqnarray}\label{vec_nl}
\vec v_{t_1 z_2} - \vec v_{t_2 z_1} + \sum_{i=1}^N (\vec v_{z_1}\cdot \nabla_{\vec x})\vec v_{z_2}  
- \sum_{i=1}^N \vec v_{z_2}\cdot \nabla_{\vec x}\vec v_{z_1}=\vec 0,  
\end{eqnarray}
where $\vec v$ is an $N$-dimensional vector and $\nabla_{\vec x}=(\partial_{x_1},..,\partial_{x_N})$.
\end{enumerate}
\end{enumerate}

Each one of the above methods of solution allows one to solve a particular class of PDEs and is not  
applicable to other classes.

Recently, several variants of the classical dressing method have been suggested, allowing to unify 
the integration algorithms for $C$ - and $S$ - integrable PDEs \cite{Z}, 
for $C$- and $Ch$- integrable PDEs \cite{ZS}, and  for $S$- and $Ch$- integrable PDEs \cite{Z5}. 
In particular, the relation between the matrix PDE (\ref{w}), integrable by the method of characteristics, and the 
$GL(M,\CC)$ SDYM equation has been recently established in \cite{Z4}. As a consequence of this result, 
it was shown that the SDYM equation admits an infinite class of lower-dimensional reductions which are 
integrable by the method of characteristics.

In this paper we extend the results of \cite{Z4}, showing 
the existence of remarkably deep relations among $S$-, $C$- and $Ch$- integrable systems. 
More precisely, we do the following. 
\begin{enumerate}
\item We show (in $\S$ 2) that the integrability properties (Lax pair, infinitely-many 
commuting symmetries, large classes of analytic solutions) of the $C$-integrable (1+1)-dimensional 
matrix B\"urgers hierarchy can be used to generate the
integrability properties of $S$-integrable PDEs in  
(2+1) dimensions, like the N-wave, KP, and DS equations; this result is achieved using a block Frobenius 
matrix reduction of the relevant matrix 
field of the matrix B\"urgers hierarchy, leading to integrable chains of matrix equations for the 
blocks of such a Frobenius matrix, followed by a systematic elimination procedure of some of these blocks.  
The construction of large classes 
of solutions of the soliton equations from solutions of the matrix B\"urgers hierarchy turns 
out to be intimately related to the construction of solutions in Sato theory \cite{Sato1,Sato2,Sato3,OSTT}.    
 On the way back, starting with the Lax pair eigenfunctions of 
the derived $S$-integrable systems, we show that the coefficients of their asymptotic expansions, 
for large values of the spectral parameter, coincide with the elements of the above integrable chains,  
obtaining an interesting spectral meaning of such chains. It follows that, 
compiling these coefficients into the Frobenius matrix, one constructs the $C$-integrable 
matrix Burgers hierarchy and its solutions from the eigenfunctions of the $S$-integrable systems. 
\item We show (in $\S$ 3) that the integrability properties of the multidimensional matrix equation (\ref{w}),   
solvable by the method of characteristics, can be used to generate the integrability properties of 
\begin{enumerate} 
\item $S$-integrable PDEs in (1+1) dimensions, like the N-wave, KdV, modified KdV (mKdV), and  
NLS equations (in \S 3.2);
\item $S$-integrable multidimensional generalizations of the $GL(M,\CC)$ SDYM equations (in \S 3.3); 
this derivation from the simpler and basic matrix equation (\ref{w}), allows one to uncover for free two 
important properties of such equations: 
a convenient parametrization, given in terms of the blocks of the Frobenius matrix, allowing one to reduce by half 
the number of equations, and the existence of a large class of solutions describing the gradient catastrophe 
of multidimensional waves.
\item $S$-integrable multidimensional Calogero equations 
\cite{Calogero_syst1,Calogero_syst2,Calogero_syst3,Calogero_syst4,ZM_Calogero} (in \S 3.4). 
\end{enumerate}
As before, these results are obtained considering a 
block Frobenius matrix reduction, leading to integrable chains, followed by a systematic elimination procedure 
of some of their elements. Vice-versa, such chains are satisfied by the coefficients of the asymptotic 
expansion, for large values of the spectral parameter, of the eigenfunctions of the soliton equations. 
\item We show (in $\S$ 4) that a proper generalization of the block Frobenius matrix reduction of the 
matrix B\"urgers hierarchy can be used to construct the integrability properties of nonlinear PDEs 
exhibiting properties in common with both $S$- and $C$- integrable equations. 
\end{enumerate}
Figure 1 below shows the diagram summarizing the connections discussed in $\S$ 2 and $\S$ 3.

We end this introduction mentioning previous work related to our main findings. i) The matrix Burgers equation (\ref{Burgers0}) 
with $B=I$, together with the block Frobenius matrix reduction (\ref{W}), have been used in \cite{AH} to construct 
some explicit solutions of the linear Schr\"odinger and diffusion equations. ii) As already mentioned, once the 
connections illustrated in \S 2 are exploited to construct 
large classes of solutions of soliton equations from simpler solutions of the matrix B\"urgers hierarchy, the 
corresponding formalism turns out to be intimately related to the construction of solutions of soliton equations in Sato theory. 

\vskip 20pt
\begin{center}
\mbox{ \epsfxsize=18cm \epsffile{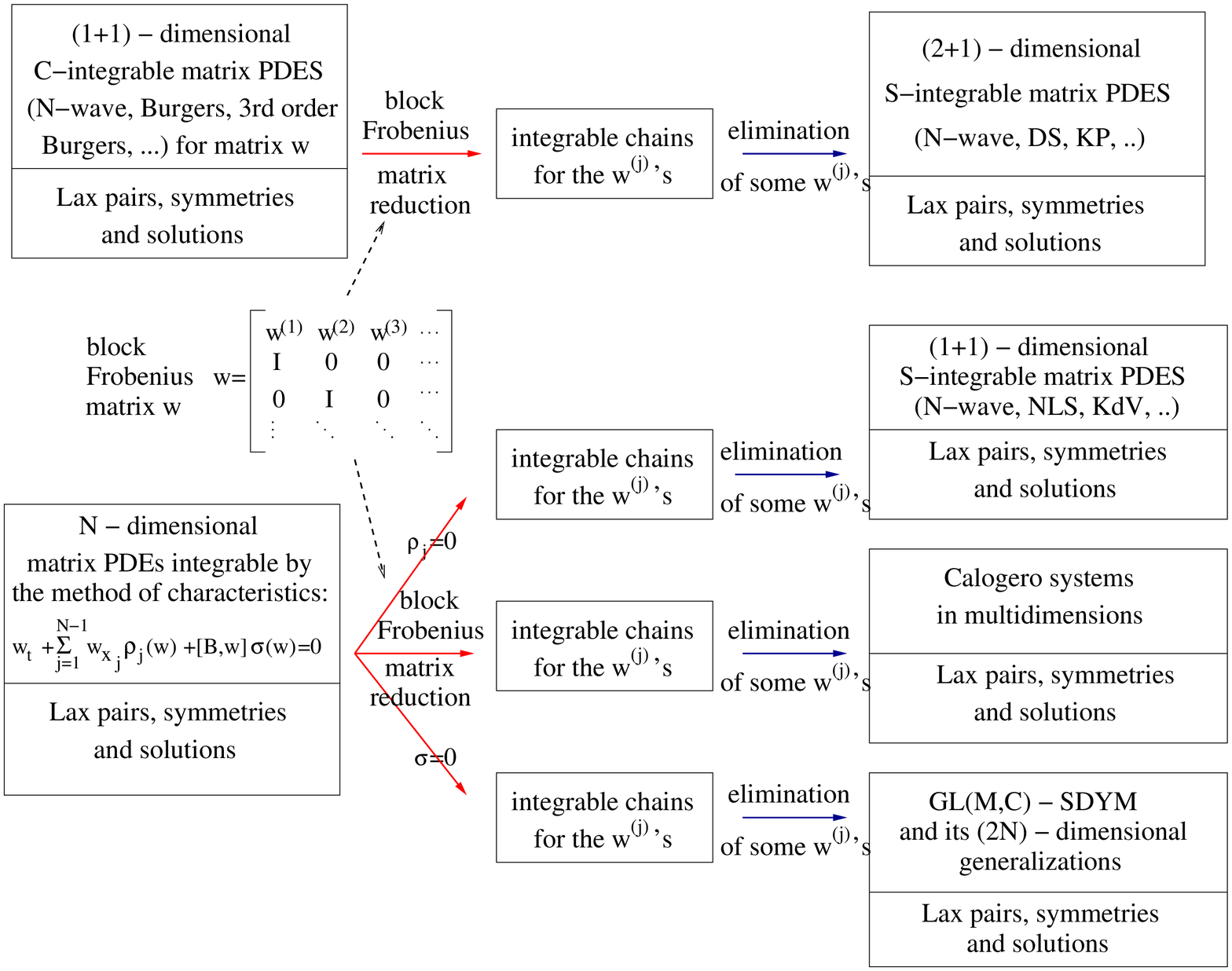}}
\end{center}
Fig. 1 The remarkable relations among PDEs integrable by 
the inverse spectral transform method, by
the method of characteristics and by the Hopf-Cole transformation.
 
\section{Relation between $C$- and $S$-integrability}
Usually $C$- and $S$- integrable systems are considered as 
completely integrable systems with different integrability features. In this section we 
show the  remarkable relations between them.

\subsection{$C$-integrable PDEs}
It is well known that the hierarchy of $C$-integrable systems associated with the matrix Hopf - Cole 
transformation 
\begin{eqnarray}\label{Hopf}
\Psi_x=w\Psi
\end{eqnarray}
can be generated by the compatibility condition  between equation (\ref{Hopf})  
and the following hierarchies of linear commuting flows (the hierarchy generated by higher $x$-derivatives 
and its replicas):
\begin{eqnarray}\label{C_flows}
&&
\Psi_{t_{nm}}=B^{(nm)}\partial^n_x\Psi,~~n,m\in\NN_+,
\end{eqnarray}
where $\Psi$ and $w$ are square matrix functions and $B^{(nm)},~n,m\in\NN_+$ are constant 
commuting square matrices. The 
integrability conditions yield the following hierarchy of $C$-integrable equations and its replicas: 
\begin{eqnarray}\label{C_hierarchy}
&&
w_{t_{nm}}+[w,B^{(nm)} W^{(n)}] - B^{(nm)} W^{(n)}_x=0,
\end{eqnarray}
where
\beq\label{def_W}
\ba{c}
W^{(n)}= W^{(n-1)}_x + W^{(n-1)} w,\;\;n\in\NN_+ ,         \\
W^{(0)}=I,\;W^{(1)}=w,\;W^{(2)}=w_x+w^2,\;W^{(3)}=w_{xx}+2w_xw+ww_x+w^3, \dots .
\ea
\eeq
and $I$ is the identity matrix.

The first three examples, together with their commuting replicas, read:

\begin{enumerate}
\item
$n=1$:
a $C$-integrable $N$-wave equation in (1+1)-dimensions:
\beq\label{n_wave}
\ba{l}
w_{t_{1m}} - B^{(1m)} w_x + [w,B^{(1m)}]w =0, 
\ea
\eeq
\item
$n=2$:
the matrix B\"urgers equation:
\begin{eqnarray}\label{Burgers}
w_{t_{2m}} - B^{(2m)} w_{xx} -2B^{(2m)} w_x w +[w,B^{(2m)}] (w_x+w^2) =0;
\end{eqnarray}
\item
$n=3$:
the 3-rd order matrix B\"urgers equation:
\beq\label{Burgers-3}
\ba{l}
w_{t_{3m}} - B^{(3m)} w_{xxx}-3 B^{(3m)}w_{xx} w  +[w,B^{(3m)}] (w_{xx}+ww_x + 2w_xw+w^3)- \\
 3B^{(3m)} w_x(w_x +  w^2)  =0.
\ea
\eeq
\end{enumerate}

The way of generating solutions of the $C$- integrable PDEs (\ref{C_hierarchy}) is elementary: take the 
general solution of equations (\ref{C_flows}):
\beq\label{sol_C_1}
\Psi(\vec{x})=\int_{\Gamma}  e^{ kx+\sum\limits_{j,m\ge 1}B^{(jm)}  t_{jm}k^j}\hat \Psi(k)d\Omega(k),
\eeq
where $\Gamma$ is an arbitrary contour in the complex $k$-plane, $\Omega(k)$ is an arbitrary measure and 
$\hat \Psi(k)$ is an arbitrary matrix function of the spectral parameter $k$, and $\vec x$ is the vector 
of all independent variables: $\vec x =\{x,t_{nm};~n,m\in \NN_+ \}$. Then 
\beq\label{sol_C_2}
w=\Psi_x\Psi^{-1}
\eeq 
solves (\ref{C_hierarchy}).

\subsection{Block Frobenius matrix structure, integrable chains and $S$-integrable PDEs}

It turns out that the $C$-integrable hierarchy of (1+1)-dimensional PDEs (\ref{C_hierarchy}), including the 
$N$-wave, B\"urgers and third order B\"urgers equations (\ref{n_wave})-(\ref{Burgers-3}) as distinguished 
examples,  
generates a corresponding hierarchy of $S$-integrable (2+1)-dimensional PDEs, including the celebrated 
$N$-wave, DS and KP equations respectively. This is possible, due to the remarkable fact that eqs. 
(\ref{Hopf}) and (\ref{C_flows}) are compatible with the following {\bf block Frobenius matrix}  
structure of the matrix function $w$:
\begin{eqnarray}\label{W}
w=\left[
\begin{array}{ccccc}
w^{(1)} & w^{(2)} & w^{(3)} &\cdots\cr
I_M & 0_M &  0_M&\cdots \cr
 0_M&I_M &  0_M&\cdots \cr
\vdots&\ddots&\ddots&\ddots
\end{array}
\right],
\end{eqnarray}
where $I_M$ and $0_M$  are the $M\times M$ identity and zero matrices, $M\in\NN_+$, and $w^{(j)},~j\in\NN_+$ are 
$M\times M$ matrix functions. This block structure of $w$ is consistent with eqs.(\ref{Hopf}) and (\ref{C_flows}) 
(and therefore with the whole $C$-integrable hierarchy (\ref{C_hierarchy})) iff 
matrix $\Psi$ is a {\bf block Wronskian matrix}:
  \begin{eqnarray}\label{Psi2}
 \Psi=\left[
\begin{array}{cccc}
 \Psi^{(11)}& \Psi^{(12)}& \Psi^{(13)}& \cdots \cr
\partial^{-1}_x\Psi^{(11)}& \partial^{-1}_x\Psi^{(12)}&
 \partial^{-1}_x\Psi^{(13)}& \cdots \cr
\partial^{-2}_x\Psi^{(11)}& \partial^{-2}_x\Psi^{(12)}&
 \partial^{-2}_x\Psi^{(13)}& \cdots \cr
\vdots&\vdots&\vdots&\ddots\cr
\end{array}
\right],
\end{eqnarray}
and
 \begin{eqnarray}\label{B}
B^{(im)}={\mbox{diag}}(\tilde B^{(im)},\tilde B^{(im)},\cdots),
 \end{eqnarray}
where the blocks $\Psi^{(ij)},~i,j\in\NN_+$ are $M\times M$ matrices, and 
$\tilde B^{(im)},\;i\in\NN_+$ are constant commuting $M\times M$ matrices. In 
equations (\ref{W})-(\ref{B}), the matrices $w$, $\Psi$ and $B^{(im)}$ are chosen to be $\infty\times \infty$ square 
matrices containing an infinite number of finite blocks; only in dealing with the construction of 
explicit solutions, it is convenient to consider a finite number of blocks.
 
Substituting the expressions (\ref{W}) and (\ref{B}) into the nonlinear PDEs
(\ref{n_wave}-\ref{Burgers-3}), one obtains the following (by construction) {\bf integrable infinite  
chains of PDEs}, for $n,m\in\NN_+$:
\begin{eqnarray}\label{n_wave_ch}
w^{(n)}_{t_{1m}} - \tilde B^{(1m)} w^{(n)}_{x}+[w^{(n+1)},\tilde B^{(1m)}] +[w^{(1)},\tilde B^{(1m)}] w^{(n)} =0,
\end{eqnarray} 
\begin{eqnarray}\label{Burgers_ch}
&&
w^{(n)}_{t_{2m}} - \tilde B^{(2m)} w^{(n)}_{xx}- 2 \tilde B^{(2m)} w^{(n+1)}_x
  -2 \tilde B^{(2m)} w^{(1)}_x w^{(n)} +\\\nonumber
  &&
  [w^{(1)},\tilde B^{(2m)}] (w^{(1)} w^{(n)} + w^{(n)}_x+ w^{(n+1)})+ 
  [w^{(2)},\tilde B^{(2m)}] w^{(n)} + [w^{(n+2)},\tilde B^{(2m)}]=0 ,
\end{eqnarray} 
\begin{eqnarray}\label{Burgers-3_ch}
&&
w^{(n)}_{t_{3m}} - \tilde B^{(3m)} \Big( w^{(n)}_{xxx}
+ 3 (w^{(1)}_{xx} w^{(n)}+ w^{(n+1)}_{xx}) + 3 w^{(1)}_x (w^{(1)} w^{(n)} + w^{(n+1)} + w^{(n)}_x)+\\\nonumber
&&
3w^{(2)}_xw^{(n)}+3w^{(n+2)}_{x}\Big) +  [w^{(1)},\tilde B^{(3m)}] \Big(w^{(n)}_{xx}+ 
2(w^{(1)}_x w^{(n)}+ w^{(n+1)}_x) +w^{(1)}( w^{(n)}_x  +         \\\nonumber
&&
w^{(1)} w^{(n)}+  w^{(n+1)})+w^{(2)}w^{(n)}+w^{(n+2)}\Big)+
[w^{(2)},\tilde B^{(3m)}] \Big(w^{(n)}_{x} +w^{(1)} w^{(n)}+ w^{(n+1)}\Big)+        \\\nonumber
&&
[w^{(3)},\tilde B^{(3m)}] w^{(n)}+[w^{(n+3)},\tilde B^{(3m)}] =0 .   
\end{eqnarray} 

From these chains, whose spectral nature will be unveiled in $\S$ 2.4, one constructs, through a systematic 
{\bf elimination of some of the blocks $w^{(j)}$}, the target $S$-integrable PDEs. Here we consider the following basic examples.

\paragraph{(2+1)-dimensional $N$-wave equation.}

 Fixing $n=1$ in eqs.(\ref{n_wave_ch}), and choosing $m=1,2$, one obtains the following complete system of equations for 
$w^{(i)}$, $i=1,2$:
\beq\label{n_wave_ch2}
\ba{l}
w^{(1)}_{t_{1m}} - \tilde B^{(1m)} w^{(1)}_{x} +[w^{(1)},\tilde B^{(1m)}] w^{(1)}+[w^{(2)},\tilde B^{(1m)}] =0,\;\;m=1,2.
\ea
\eeq 
Eliminating $w^{(2)}$ from  equations (\ref{n_wave_ch2})  one obtains the classical (2+1)-dimensional $S$-integrable 
$N$-wave equation:
\begin{eqnarray}\label{S_n_w}
&&
[w^{(1)}_{t_{11}},\tilde B^{(12)}]-
[w^{(1)}_{t_{12}},\tilde B^{(11)}]  - \tilde B^{(11)} w^{(1)}_{x}\tilde B^{(12)} +
\tilde B^{(12)} w^{(1)}_{x}\tilde B^{(11)} +\\\nonumber
&&
[[w^{(1)},\tilde B^{(11)}],[ w^{(1)},\tilde B^{(12)}]]=0.
\end{eqnarray}

\paragraph{DS-type equation.}
Choosing $n=1,2$ in equations (\ref{n_wave_ch}),  $n=1$ in eq.(\ref{Burgers_ch}), $m=1$ in both equations, and simplifying 
the notation as follows:
\begin{eqnarray}\label{red_var}
t_{j}=t_{j1},\;\;\tilde B^{(j)}=\tilde B^{(j1)},\;\;j\in\NN_+,
\end{eqnarray}  
one obtains the following complete system of equations for $w^{(i)}$, $i=1,2,3$:
\begin{eqnarray}\label{n_wave_ch31}\label{DS_1}
&&
w^{(1)}_{t_{1}} - \tilde B^{(1)} w^{(1)}_{x}+[w^{(2)},\tilde B^{(1)}] +[w^{(1)},\tilde B^{(1)}] w^{(1)} =0,
\\\label{n_wave_ch32}\label{DS_2}
&&
w^{(2)}_{t_1} - \tilde B^{(1)} w^{(2)}_{x}+[w^{(3)},\tilde B^{(1)}] +[w^{(1)},\tilde B^{(1)}] w^{(2)} =0,\\
\label{Burgers_ch2}\label{DS_3}
&&
w^{(1)}_{t_2} - \tilde B^{(2)} w^{(1)}_{xx}- 2 \tilde B^{(2)} w^{(2)}_x -2 \tilde B^{(2)} w^{(1)}_x w^{(1)} +\\\nonumber
&&
[w^{(1)},\tilde B^{(2)}] (w^{(1)} w^{(1)} + w^{(1)}_x+ w^{(2)})+ 
[w^{(2)},\tilde B^{(2)}] w^{(1)} + [w^{(3)},\tilde B^{(2)}]=0.
\end{eqnarray} 
Using eqs.(\ref{n_wave_ch31}) and (\ref{n_wave_ch32}), one can eliminate $w^{(3)}$ and $w^{(2)}$ from  
eq.(\ref{Burgers_ch2}). 
In the  case $\tilde B^{(2)}=\alpha \tilde B^{(1)}$ ($\alpha$ is a scalar),  this results in 
the following equation for $w^{(1)}$:
\begin{eqnarray}\label{off}
&&
[w^{(1)}_{t_2} , \tilde B^{(1)}] + \alpha\Big(w^{(1)}_{t_1t_1} - \tilde B^{(1)} w^{(1)}_{xx} \tilde B^{(1)} + 
[[w^{(1)},\tilde B^{(1)}],w^{(1)}_{t_1}] +\\\nonumber
&&
B^{(1)} w_x [ B^{(1)},w] - 
[B^{(1)},w] w_x B^{(1)}\Big)=0.
\end{eqnarray}
In the simplest case of square matrices ($M=2$), with $\tilde B^{(1)}=\beta \; {\mbox{diag}}(1,-1)$ ($\beta$ is a scalar constant), 
this equation reduces to the DS system:
\begin{eqnarray}\label{DS}
&&
\tilde\beta q_{t_2} -\frac{1}{2}( q_{xx} + \frac{1}{\beta^2}q_{t_1t_1}) - 2   (\varphi q +2 r q^2)=0,
\\\nonumber
&&
-\tilde\beta r_{t_2} -\frac{1}{2}( r_{xx} +\frac{1}{\beta^2} r_{t_1t_1}) - 2  (\varphi r +2 q r^2)=0,
\\\nonumber
&&
 \varphi_{xx}-\frac{1}{\beta^2}\varphi_{t_1t_1} + 4  (rq)_{xx} =0,
\end{eqnarray}
where
\begin{eqnarray}
&&
q=w^{(1)}_{12},\;\; r=w^{(1)}_{21},\;\; \varphi=(w^{(1)}_{11}+w^{(1)}_{22})_x , \;\; \tilde\beta=\frac{1}{\alpha\beta}.
\end{eqnarray}
If $\tilde \beta=i$, this system admits the reduction $r=\bar q$:
\begin{eqnarray}
&&
i q_{t_2} -\frac{1}{2}( q_{xx} + \frac{1}{\beta^2}q_{t_1tu_1}) - 
2   (\varphi q +2 \bar q q^2)=0,
\\\nonumber
&&
 \varphi_{xx}-\frac{1}{\beta^2}\varphi_{t_1t_1} + 4  (\bar q q)_{xx} =0,
 \end{eqnarray}
becoming DS-I and DS-II if $\beta^2=-1$ and $\beta^2=1$ respectively.

\paragraph{KP.}

To derive the celebrated KP equation, choose $M=1$,  
take eqs.(\ref{Burgers_ch}) with $n=1,2$, and eq. (\ref{Burgers-3_ch}) with $n=1$, $\tilde B^{(2)}=\beta$, 
$\tilde B^{(3)}=-1$, where $\beta$ is a scalar parameter, obtaining: 
\begin{eqnarray}\label{KP_1}
&&
w^{(1)}_{t_2} -\beta \Big(w^{(1)}_{xx} +2 w^{(1)} w^{(1)}_{x}+ 2 w^{(2)}_{x}\Big)=0
,\\\nonumber
&&
w^{(2)}_{t_2} - \beta \Big(w^{(2)}_{xx} +2 w^{(2)} w^{(1)}_{x}+
 2 w^{(3)}_{x}\Big)=0
,\\\nonumber
&&
w^{(1)}_{t_3} +w^{(1)}_{xxx} +3 \Big((w^{(1)})^2 w^{(1)}_x + 
(w^{(1)}_x)^2+ w^{(1)} w^{(1)}_{xx}\Big) +
\\\nonumber
&&
3 \Big( w^{(2)}_{xx} + 
w^{(2)} w^{(1)}_x+w^{(1)} w^{(2)}_x\Big)+3 w^{(3)}_x =0,
\end{eqnarray} 
where we have set $m=1$ and used again the notations (\ref{red_var}).

After eliminating $w^{(2)}$ and $w^{(3)}$, one obtains the scalar potential KP for $u=w^{(1)}$, $y=t_2$, $t=t_3$:
 \begin{eqnarray}\label{KP}
 \Big(
 u_{t} +\frac{1}{4} u_{xxx} +\frac{3}{2}u_x^2
 \Big)_x+\frac{3}{4\beta^2} u_{yy}=0 .
 \end{eqnarray}
KP-I and KP-II correspond to $\beta^2=-1$ and $\beta^2=1$ respectively.

\subsection{Lax pairs for the $S$-integrable systems}
\label{Section:S-C_sp}
Also the Lax pairs for the $S$-integrable systems derived in \S 2.2 can be constructed in a similar way, from the system 
(\ref{Hopf}), (\ref{C_flows}).
We first observe that, due to equation (\ref{Hopf}), equations (\ref{C_flows}) can be rewritten as
\beq\label{t12_m}
\ba{l}\Psi_{t_{nm}}=B^{(nm)}W^{(n)} \Psi.
\ea
\eeq
Due to the block Frobenius structure of $w$, 
it is convenient to work with the duals of equations (\ref{Hopf}) and (\ref{t12_m}):
\begin{eqnarray}\label{x_mm}
&&
\tilde \Psi_{x}=-\tilde \Psi w,\\\label{t12_mm}
\label{t13_mm}
&&
\tilde \Psi_{t_{nm}}=-\tilde \Psi  B^{(nm)} W^{(n)} .
\end{eqnarray}
Substituting (\ref{W}) and (\ref{B}) into equations (\ref{x_mm}-\ref{t13_mm}),  
one obtains a system of linear chains for the blocks of matrix $\tilde \Psi$. The first few equations 
involving the blocks of the first row read:
\begin{eqnarray}\label{sp_x}
&&
\tilde \Psi^{(1n)}_{x}+\tilde \Psi^{(11)}w^{(n)} + \tilde \Psi^{(1(n+1))}=0, \\\label{sp_t3} 
&&
\tilde  \Psi^{(1n)}_{t_{1m}}+ \tilde \Psi^{(11)}  \tilde B^{(1m)} w^{(n)} + \tilde \Psi^{(1(n+1))} \tilde B^{(1m)}=0, \\
\label{sp_t12}
\label{sp_t4} 
 && 
\tilde \Psi^{(1n)}_{t_{2m}}+ \tilde \Psi^{(11)} \tilde B^{(2m)} (w^{(n)}_x+w^{(1)}w^{(n)}+w^{(n+1)}) + 
 \tilde  \Psi^{(12)} \tilde B^{(2m)} w^{(n)} + \tilde  \Psi^{(1(n+2))}\tilde B^{(2m)} = 0,  \\ \label{sp_t5}
&& 
\tilde \Psi^{(1n)}_{t_{3m}} +\Big(\tilde \Psi^{(11)} \tilde B^{(3m)} 
\Big(w^{(n)}_{xx} + 
2(w^{(1)}_x w^{(n)}+ w^{(n+1)}_x) +w^{(1)}(w^{(n)}_x + w^{(1)} w^{(n)}+w^{(n+1)})+\\\nonumber
&&
w^{(2)}w^{(n)}+w^{(n+2)}\Big)
+
\tilde \Psi^{(12)} \tilde B^{(3m)} \Big(w^{(n)}_{x} +w^{(1)} w^{(n)}+ w^{(n+1)}\Big)
+ \tilde \Psi^{(13)} \tilde B^{(3m)} w^{(n)}+ \\\nonumber
&&
 \tilde \Psi^{(1(n+3))} \tilde B^{(3m)} = 0,
\end{eqnarray}
where $n\in\NN_+$ and $\tilde \Psi^{(ij)}$ is the ($i,j$)-block of matrix $\tilde \Psi$. 

\paragraph{Lax pair for the $N$-wave equation.}
Setting $n=1$ into eqs.(\ref{sp_x},\ref{sp_t3}) and eliminating $\tilde \Psi^{(12)}$, one obtains 
(the dual of) the Lax pair for the $N$-wave equation (\ref{S_n_w}):
\beq\label{sp2_t12_d}
\ba{l}
{\tilde\psi}_{t_{1m}}-{\tilde\psi}_{x}\tilde B^{(1m)}+ {\tilde\psi} [\tilde B^{(1m)}, w^{(1)}] =0,\;\;m=1,2,
\ea
\eeq
where $\tilde\psi=\tilde\Psi^{(11)}$.
The dual of it, is the well-known Lax pair of the $N$-wave equation (\ref{S_n_w}):
\beq\label{sp2_t12_dd}
\ba{l}
{\psi}_{t_{1m}}-\tilde B^{(1m)} {\psi}_{x}- [\tilde B^{(1m)}, w^{(1)}] {\psi} =0,\;\;m=1,2.
\ea
\eeq 
Of course, the compatibility 
condition of eqs.(\ref{sp2_t12_d}) and/or eqs. (\ref{sp2_t12_dd}) yields the nonlinear system (\ref{S_n_w}).

\paragraph{Lax pair for DS.}
In this paragraph we set $m=1$ in the integrable chains, and use the notation (\ref{red_var}). 
The first equation of the dual of the Lax pair is eq. (\ref{sp2_t12_d}) with $m=1$. To derive the second equation, we set $m=n=1$ into 
eq. (\ref{sp_t4}), and eliminate the fields $\tilde \Psi^{(12)}$, $\tilde\Psi^{(13)}$, using eq. (\ref{sp_x}) with $n=1,\; 2$. 
In this way  one obtains the dual of the Lax pair for DS-type equations:
 \begin{eqnarray}
\label{sp2_t3_d} 
  &&
{\tilde\psi}_{t_1}-{\tilde\psi}_{x}\tilde B^{(1)}+ {\tilde\psi} [\tilde B^{(1)}, w^{(1)}] =0,  \\\nonumber
&&
 \tilde\psi_{t_2}+\tilde\psi_{xx}\tilde B^{(2)}+\tilde\psi_x 
 [ w^{(1)},\tilde B^{(2)}] + \tilde\psi \tilde s=0,\;\;\\\nonumber
&&
 \tilde s= [\tilde B^{(2)},w^{(2)}] +w^{(1)}_x \tilde B^{(2)} + 
 \tilde B^{(2)} w^{(1)}_x +
 [\tilde B^{(2)} , w^{(1)}] w^{(1)} .
\end{eqnarray} 
Therefore the Lax pair reads:
\beq\label{sp2_t3_dd}
\ba{l}
{\psi}_{t_1}-\tilde B^{(1)}{\psi}_{x}-[\tilde B^{(1)}, w^{(1)}]{\psi}  =0,         \\ 
\psi_{t_2}-\tilde B^{(2)}\psi_{xx}+[ w^{(1)},\tilde B^{(2)}]\psi_x -  s(y)\psi=0,\\
  s= [\tilde B^{(2)},w^{(2)}] +2  \tilde B^{(2)}w^{(1)}_x +
 [\tilde B^{(2)} , w^{(1)}] w^{(1)} .
\ea
\eeq
The compatibility conditions of equations (\ref{sp2_t3_d}) or (\ref{sp2_t3_dd}) yield a nonlinear system equivalent 
to the system (\ref{DS_1}-\ref{DS_3}).

\paragraph{Lax pair for  KP.}
In this paragraph we use the notations (\ref{red_var}) as well.
The first equations of the dual of the Lax pair for  KP are the scalar versions of eq.(\ref{sp2_t3_d}b) and eq.(\ref{sp2_t3_dd}b) 
respectively, with $\tilde B^{(2)}=\beta$ and $\tilde B^{(3)}=-1$.
To write the second equation of the Lax pair for  KP, we must take the scalar version of eq.(\ref{sp_t5}) with $m=n=1$,  
and eliminate  $\tilde \Psi^{(1i)}$, $i=2,3,4$ using eq.s(\ref{sp_t4}) for $n=1,2$. As a result, the dual of the Lax 
pair reads 
  \begin{eqnarray}\label{sp_KP_d}
  &&
 \frac{1}{\beta} \tilde\psi_{t_2} + \tilde\psi_{xx} + 2 \tilde\psi u_x=0,\\\nonumber
  &&
  \tilde\psi_{t_3} +\tilde\psi_{xxx} + 3 \tilde\psi_x u_x - \frac{3}{2} \tilde\psi\left(\frac{u_{t_2}}{\beta} - u_{xx}\right)=0,
  \end{eqnarray} 
and  the  Lax pair is
  \begin{eqnarray}\label{sp_KP_dd}
  &&
  \frac{1}{\beta}\psi_{t_2} - \psi_{xx} - 2  u_x \psi = 0,\\\nonumber
  &&
  \psi_{t_3} +\psi_{xxx} + 3  u_x \psi_x+ \frac{3}{2} \left(\frac{u_{t_2}}{\beta} + u_{xx}\right)\psi=0.
  \end{eqnarray}

\subsection{From the Lax pairs of $S$-integrable PDEs to $C$-integrable PDEs}
\label{Section:dual_n_wave}

As usual in the IST for (2+1)-dimensional soliton equations, one introduces the spectral parameter $\lambda$ 
into the Lax pairs (\ref{sp2_t12_dd}),(\ref{sp2_t3_dd}) and (\ref{sp_KP_dd}) as follows   
\begin{eqnarray}\label{UV}
\psi(\lambda;\vec{x}) =  \chi (\lambda;\vec{x})e^{\lambda x I+\sum\limits_{i,m\ge 1}
 \tilde B^{(im)} t_{im}\lambda^i},
\end{eqnarray} 
obtaining, respectively, the following spectral systems for the new eigenfunction $\chi$:
\begin{eqnarray}
\label{sp2_t12_V}
&& 
 \chi_{t_{1m}}-\tilde B^{(1m)}\chi_{x}- \lambda[\tilde B^{(1m)},\chi]-[\tilde B^{(1m)} w^{(1)}]  \chi =0,
 \end{eqnarray}
 \begin{eqnarray}\label{sp2_t12_V2}
 &&
\chi_{t_1}-\tilde B^{(1)}\chi_{x}- \lambda[\tilde B^{(1)},\chi]-
[\tilde B^{(1)} w^{(1)}] \chi ,         \\\label{sp2_t3_V} 
  &&
  \chi_{t_2}-\tilde B^{(2)}\chi_{xx}+\lambda^2 [\chi,\tilde B^{(2)}]  - 
  2\lambda \tilde  B^{(2)} \chi_x
  +[ w^{(1)},\tilde B^{(2)}](\chi_x+\lambda \chi)  - s\chi =0,\\\nonumber
  &&
   s= [\tilde B^{(2)},w^{(2)}] +2  \tilde B^{(2)}w^{(1)}_x +
 [\tilde B^{(2)} , w^{(1)}] w^{(1)}.   
\end{eqnarray} 
\begin{eqnarray}\label{sp2_t3_V_KP} 
  &&
  \chi_{t_2}-\beta\Big(2\lambda \chi_x  +\chi_{xx} + 2 \chi w_x\Big)=0,\\\nonumber
  &&
 \chi_{t_3}+
 3\lambda \chi_{xx} +3\lambda^2  \chi_{x} + \chi_{xxx}+3\lambda \chi w_x + 
 \frac{3}{2} \chi \Big(\frac{w_{t_2}}{\beta} + w_{xx}\Big)=0.
\end{eqnarray} 

It is now easy to verify that the coefficients of the $\lambda$ - large expansion of 
the eigenfunction $\chi$ satisfy the infinite chains (\ref{n_wave_ch}-\ref{Burgers-3_ch}):
\begin{eqnarray}\label{V_inf}
\chi(\lambda;\vec{x}) = I - \sum_{n\ge 1}\frac{ w^{(n)}(\vec{x})}{\lambda^i}. 
\end{eqnarray}
Therefore we have obtained the {\bf spectral interpretation} of such chains. In addition, since the infinite chains 
(\ref{n_wave_ch}-\ref{Burgers-3_ch}) for the $w^{(n)}$'s 
are equivalent, via the Frobenius structure (\ref{W}), to $C$-integrable systems, we have also shown how to go backward,  
from $S$- to $C$- integrability.

\subsection{Construction of solutions and Sato theory}
 \label{Section:S-C_sol}
In order to construct solutions of the S-integrable PDEs generated in \S 2.2 from the elementary solution scheme 
(\ref{sol_C_1}),(\ref{sol_C_2}) of the matrix Burgers hierarchy, we consider the matrices $w$ and $\Psi$ to be 
finite matrices consisting of $n_0\times n_0$ blocks (this can be done assuming that $\Psi^{(1(n_0+1))}=0$),   
where $n_0$ is an arbitrary positive integer greater than the number of blocks ${w^{(j)}}'s$ involved in the S-integrable 
PDE under consideration. Taking into account the structures of $w$ and $\Psi$ given by 
eqs.(\ref{W}) and (\ref{Psi2}) respectively, we have that
\beq\label{W_fin}
w=\left[ 
\ba{ccccc}
w^{(1)} & w^{(2)} & \cdots & \cdots & w^{(n_0)} \cr
I_M     &  0_M    & \cdots &\cdots &   0_M      \cr
0_M     & \ddots  & \ddots &\ddots & \vdots     \cr
\vdots  & \ddots  &  \ddots& \ddots  & \vdots     \cr
\vdots  & \ddots  &   0_M  &  I_M  &  0_M       
\ea
\right], \Psi=\left[ 
\ba{cccc}
\Psi^{(11)}& \Psi^{(12)}& \cdots & \Psi^{(1n_0)} \cr
\partial_x^{-1}\Psi^{(11)}& \partial_x^{-1}\Psi^{(12)}& \cdots & \partial_x^{-1}\Psi^{(1n_0)} \cr 
\vdots  & \vdots  &  \vdots& \vdots      \cr
\partial_x^{-n_0+1}\Psi^{(11)}& \partial_x^{-n_0+1}\Psi^{(12)}& \cdots & \partial_x^{-n_0+1}\Psi^{(1n_0)}       
\ea
\right]. 
\eeq
We remark that the $n_0$ blocks  $\Psi^{(1j)}$, $j=1,\dots,n_0$ of $\Psi$ are defined, via (\ref{sol_C_1}), 
by equations 
 \begin{eqnarray}
\Psi^{(1j)}(\vec x) = \int \;e^{k x + 
 \sum\limits_{i,m\ge 1}\tilde B^{(im)} t_{im}k^{i}}\hat\Psi^{(1j)}(k)d\Omega (k) ,\;\;j=1,\dots,n_0
\end{eqnarray}
in terms of the arbitrary spectral functions $\hat\Psi^{(1j)}$, while the remaining blocks are constructed 
through the equations $\Psi^{(ij)}=\partial_x^{i-1}\Psi^{(1j)}$. Then, via (\ref{sol_C_2}), the 
components of the $M\times M$ blocks $w^{(i)}$ are expressed in terms of $\Psi$ through the compact formula
 \begin{eqnarray}\label{sol_S}
 w^{(i)}_{\alpha\beta}=
 (\Psi_x\Psi^{-1})_{\alpha(iM-M+\beta)},\;\;\alpha,\beta=1\dots,M,~~i=1,\dots,n_0.
 \end{eqnarray}

This formula is intimately connected to those obtained via Sato theory. To see it, we consider the simplest case 
of scalar blocks ($M=1$), containing the example of the KP equation. Then equation (\ref{sol_S}) becomes 
\beq
w^{(i)}=\frac{\triangle^{(i)}}{\triangle},
\eeq
where
\beq
\triangle=\left| 
\ba{cccc}
\partial_x^{n_0-1}f^{(1)} & \partial_x^{n_0-1}f^{(2)} & \cdots & \partial_x^{n_0-1}f^{(n_0)} \cr
\partial_x^{n_0-2}f^{(1)} & \partial_x^{n_0-2}f^{(2)} & \cdots & \partial_x^{n_0-2}f^{(n_0)} \cr
\vdots                    & \vdots                    & \vdots & \vdots                      \cr
f^{(1)}                    & f^{(2)}                   & \cdots &  f^{(n_0)}
\ea
\right|, 
\eeq
\beq
\triangle^{(i)}=\left| 
\ba{ccccccc}
\partial_x^{n_0-1}f^{(1)} & \partial_x^{n_0-1}f^{(2)} & \cdots & \cdots & \cdots & \cdots & \partial_x^{n_0-1}f^{(n_0)} \cr
\vdots                    & \vdots                    & \vdots & \vdots & \vdots & \vdots  &    \vdots                    \cr
\partial_x^{n_0-i}f^{(1)} & \partial_x^{n_0-i}f^{(2)} & \cdots & \cdots & \cdots & \cdots & \partial_x^{n_0-i}f^{(n_0)} \cr
\partial_x^{n_0}f^{(1)} & \partial_x^{n_0}f^{(2)} & \cdots & \cdots & \cdots & \cdots & \partial_x^{n_0}f^{(n_0)} \cr
\partial_x^{n_0-i-2}f^{(1)} & \partial_x^{n_0-i-2}f^{(2)} & \cdots & \cdots & \cdots & \cdots & \partial_x^{n_0-i-2}f^{(n_0)} \cr
\vdots                    & \vdots                    & \vdots & \vdots& \vdots& \vdots& \vdots                      \cr
f^{(1)}                    & f^{(2)}                   & \cdots& \cdots& \cdots& \cdots &  f^{(n_0)}
\ea
\right|, 
\eeq
where $f^{(j)}=\partial_x^{-(n_0-1)}\Psi^{(1j)},~j=1,\dots,n_0$, equivalent to the formula obtained using Sato theory \cite{OSTT}.

\section{Relation between $Ch$- and $S$-integrability}

Following the same strategy illustrated in \S 2, in this section we establish the deep relations between 
the matrix PDE (\ref{w}), recently introduced in \cite{SZ} 
and  integrated there by the method of characteristics, and i) (1+1)-dimensional  $S$-integrable 
soliton equations like the KdV and NLS equations; ii) the $GL(M,\CC )$ - SDYM equation and its 
multidimensional generalizations; iii) the multidimensional Calogero systems 
\cite{Calogero_syst1,Calogero_syst2,Calogero_syst3,Calogero_syst4,ZM_Calogero}. 
In this section, matrix $w$ must be diagonalizable. 

\subsection{Matrix equations integrable by the method of characteristics}

Consider the following matrix eigenvalue problem
\beq\label{H_sp}
w(\vec{x})\Psi(\Lambda;\vec{x})= \Psi(\Lambda;\vec{x})\Lambda(\vec{x}),
\eeq
for the matrix $w(\vec{x})$, where $\Lambda (\vec{x})$ is the diagonal matrix of eigenvalues, $\Psi(\Lambda;\vec{x})$ 
is a suitably normalized matrix of eigenvectors, and associate with it   
 the following flows for $\Psi(\Lambda;\vec{x})$:  
\beq\label{H_flows}
\ba{l}
\Psi_{t_{mk}}+\sum_{j=1}^N \Psi_{x_{jk}}\rho^{(mjk)}(\Lambda) - B^{(mk)} \Psi\sigma^{(mk)}(\Lambda)=0,\;\;\; m\in\NN_+,\;\;k=1,2 ,
\ea
\eeq
where $B^{(mk)}$ are constant commuting matrices as in $\S$ 2 and $\vec{x}$ is the vector of all independent variables: 
$\vec{x}=(x_{jk},t_{nm},\;j,n,m\in\NN_+,k=1,2)$.   
The compatibility between the flows (\ref{H_flows}) implies the following commuting quasilinear PDEs for the eigenvalues: 
\beq\label{lambda_flows}
\ba{l}
\Lambda_{t_{mk}} + \sum_{j=1}^N \Lambda_{x_{jk}} \rho^{(mjk)}(\Lambda) =0,~~~m\in\NN_+, ~~k=1,2; 
\ea
\eeq
the additional compatibility with the eigenvalue problem (\ref{H_sp}) implies the following nonlinear PDEs:
\beq\label{w_m}
\ba{l}
w_{t_{mk}} + \sum\limits_{i=1}^N w_{x_{ik}}\rho^{(mik)}(w) + [w,B^{(mk)}]\sigma^{(mk)}(w)=0,~~~m\in\NN_+,\;\;k=1,2,
\ea
\eeq
commuting replicas of equation (\ref{w}).

The way of solving equations (\ref{w_m}) is as follows \cite{SZ}. Consider the general solution 
of equations (\ref{lambda_flows}) with $k=1,2$, characterized by the following non-differential equations: 
\begin{eqnarray}\label{Lambda}
\Lambda &=& E\left(x_{11}I-\sum_{m\ge1}\rho^{(m11)}(\Lambda)t_{m1},\dots ,x_{N1}I-\sum_{m\ge1}\rho^{(mN1)}(\Lambda)t_{m1};\right. 
\\\nonumber
&&\left.x_{12}I-\sum_{m\ge1}\rho^{(m12)}(\Lambda)t_{m2},\dots ,x_{N2}I-\sum_{m\ge1}\rho^{(mN2)}(\Lambda)t_{m2}
\right),
\end{eqnarray}
where $E$ is an arbitrary diagonal matrix function of $2N$ arguments; 
then the general solution of the linear matrix PDEs (\ref{H_flows}) 
for $\Psi$, with the convenient parametrization $\Psi_{ii}=1$, is given by
\begin{eqnarray}\label{w_eigP_expl}
\Psi_{\alpha\beta}&=& F_{\alpha\beta} \left(x_{11}-
\sum_{m\ge 1}\rho^{(m11)}(\Lambda_\beta)t_{m1},\dots ,x_{N1}-\sum_{m\ge 1}\rho^{(mN1)}(\Lambda_\beta)t_{m1};\right. \\\nonumber
&&\left.x_{12}I-\sum_{m\ge 1}\rho^{(m12)}(\Lambda_\beta)t_{m2},\dots ,x_{N2}-\sum_{m\ge 1}\rho^{(mN2)}(\Lambda_\beta)t_{m2}
\right)\times
 \\\nonumber
&&
e^{\sum\limits_{k=1}^2\sum\limits_{m\ge 1} 
B^{(mk)}_{\alpha} \sigma^{(mk)}(\Lambda_\beta) t_{mk}\Big)}
,\;\;\alpha,\beta=1,2,\dots
\end{eqnarray}
where  $F_{\alpha\beta}$ are arbitrary scalar functions of $2 N$ arguments, with $F_{\alpha\alpha}=1$. Then 
\beq\label{sol_Ch}
w=\Psi\Lambda\Psi^{-1}
\eeq
solves the nonlinear PDEs (\ref{w_m}).

Now we proceed as in $\S$2, assuming for $w$ the same block Frobenius matrix structure (\ref{W}), consistent with 
equations (\ref{H_sp}) and (\ref{H_flows}) (and then with the hierarchies (\ref{w_m})) iff the matrices $B^{(mk)}$ 
are given as in (\ref{B}), the diagonal matrix of eigenvalues $\Lambda(\vec x)$ has the block-structure
\begin{eqnarray}
\Lambda(\vec x)={\mbox{diag}}[\Lambda^{(1)}(\vec x),\;\;\Lambda^{(2)}(\vec x),\cdots],
\end{eqnarray}
and  
\begin{eqnarray}\label{Psi13}
\Psi=\left[\begin{array}{cccc}
\Psi^{(11)} & \Psi^{(22)} \Lambda^{(2)} & \Psi^{(33)} {\Lambda^{(3)}}^2                            &\cdots  \cr
\Psi^{(11)}(\Lambda^{(1)})^{-1} & \Psi^{(22)} & \Psi^{(33)} \Lambda^{(3)}                          & \cdots \cr
\Psi^{(11)}(\Lambda^{(1)})^{-2} & \Psi^{(22)}(\Lambda^{(2)})^{-1}&\Psi^{(33)}                      &\cdots  \cr
\Psi^{(11)}(\Lambda^{(1)})^{-3} &  \Psi^{(22)}(\Lambda^{(2)})^{-2}&\Psi^{(33)}(\Lambda^{(3)})^{-1} &\cdots  \cr
\vdots                          & \vdots                          & \vdots                         &\ddots
\end{array}\right]
\end{eqnarray}

Using the same strategy as in $\S$ 2, we show that,   
\begin{enumerate}
 \item if $\rho^{(mjk)}=0$, $k=1,2$, implying $\Lambda=const$, equations (\ref{w_m}) generate classical 
(1+1)-dimensional $S$-integrable PDEs like the $N$-wave, NLS, KdV and mKdV equations;
 \item if $B^{(mk)}=0$ (or $\sigma^{(mk)}=0$), $k=1,2$, equations (\ref{w_m}) generate the 
$GL(M,\CC )$-SDYM equation and its ($2N+2$) - dimensional generalization;
 \item if $B^{(m1)}=0$ (or $\sigma^{(m1)}=0$) and  
$ \rho^{(mi2)}=0$, equations (\ref{w_m}) generate  Calogero systems.
\end{enumerate} 

\subsection{Derivation of (1+1)-dimensional $S$-integrable PDEs}
\label{Section:1+1}
Let $\rho^{(mjk)}=0$ in (\ref{H_flows}-\ref{w_m}), implying $\Lambda=const$ (isospectral flows), and  
let $\sigma^{(mk)}(\Lambda)=\Lambda^m$, i.e.:
\begin{eqnarray}\label{sp_11}
&&
w\Psi= \Psi \Lambda,\\\label{sp_tau_m}
\label{sp_t_m}
&&
\Psi_{t_{mk}} - B^{(mk)} \Psi \Lambda^m=0, \;\;\; m,k\in\NN_+. 
\end{eqnarray}

The compatibility conditions for the system (\ref{sp_11}-\ref{sp_t_m}) yields, for $m\in\NN_+$:
\begin{eqnarray}\label{wS1}
\label{wS2}
&&
w_{t_{mk}}  + [w,B^{(mk)}]w^m=0.
\end{eqnarray}
We remark that these equations are equivalent to eqs.(\ref{C_hierarchy})
with $\partial^j_x w=0$, $\forall j$. Consequently, the discrete chains generated by the eq.(\ref{wS1}), with 
$m=1$, $k=1,2$ and  $m=2,3$, $k=1$,  are given by the eqs.(\ref{n_wave_ch}-\ref{Burgers-3_ch}) 
with $\partial^j_x w=0$, $\forall j$:
\begin{eqnarray}\label{n_wave_ch_11}
\label{n_wave_ch_111}
&&
w^{(n)}_{t_{1k}} +[w^{(n+1)},\tilde B^{(1k)}] +[w^{(1)},\tilde B^{(1k)}] w^{(n)} =0,\;\;k=1,2,
\end{eqnarray}
\begin{eqnarray}\label{Burgers_ch_11}
&&
w^{(n)}_{t_2} 
   +
  [w^{(1)},\tilde B^{(2)}] (w^{(1)} w^{(n)} +  w^{(n+1)})+ 
  [w^{(2)},\tilde B^{(2)}] w^{(n)} + [w^{(n+2)},\tilde B^{(2)}]=0 ,
\end{eqnarray} 
\begin{eqnarray}\label{Burgers-3_ch_11}
&&
w^{(n)}_{t_3} +  [w^{(1)},\tilde B^{(3)}] \Big( w^{(1)} (      
w^{(1)} w^{(n)}+  w^{(n+1)})+w^{(2)}w^{(n)}+w^{(n+2)}\Big)+ \\\nonumber
&&
[w^{(2)},\tilde B^{(3)}] \Big(w^{(1)} w^{(n)}+ w^{(n+1)}\Big)+       
[w^{(3)},\tilde B^{(3)}] w^{(n)}+[w^{(n+3)},\tilde B^{(3)}] =0 ,  
\end{eqnarray} 
for $n\in\NN_+$, where, in equations (\ref{Burgers_ch_11}),(\ref{Burgers-3_ch_11}), we have used the simplifying notation 
(\ref{red_var}).
\paragraph{(1+1)-dimensional $N$-wave equation.} 
Setting $n=1$ in eqs. (\ref{n_wave_ch_11}), and eliminating $w^{(2)}$, one obtains the well-known 
$S$- integrable $N$-wave system in (1+1)-dimensions:
\begin{eqnarray}\label{S_n_w_11}
[w^{(1)}_{t_{11}},\tilde B^{(12)}]-[w^{(1)}_{t_{12}},\tilde B^{(11)}]  +
[[w^{(1)},\tilde B^{(11)}],[ w^{(1)},\tilde B^{(12)}]]=0.
\end{eqnarray}

\paragraph{NLS.} 
Eq.(\ref{n_wave_ch_11}) with $k=1$, $n=1,2$ and eq.(\ref{Burgers_ch_11}) with $n=1$ are a complete system of PDEs for 
$w^{(j)}$, $j=1,2,3$:
\begin{eqnarray}\label{S_11}
&&
w^{(1)}_{t_{1}} +[w^{(2)},\tilde B^{(1)}] +[w^{(1)},\tilde B^{(1)}] w^{(1)} =0,\\\nonumber
\label{S_12}&&
w^{(2)}_{t_{1}} +[w^{(3)},\tilde B^{(1)}] +[w^{(1)},\tilde B^{(1)}] w^{(2)} =0,\\\nonumber
&&
w^{(1)}_{t_2} 
  +
  [w^{(1)},\tilde B^{(2)}] (w^{(1)} w^{(1)} + w^{(2)})+ 
  [w^{(2)},\tilde B^{(2)}] w^{(1)} + [w^{(3)},\tilde B^{(2)}]=0
\end{eqnarray} 
In the case $\tilde B^{(2)}=\alpha \tilde B^{(1)}$ ($\alpha$ is a scalar constant)  this system results in 
the following equation for $w^{(1)}$:
\begin{eqnarray}\label{off_11}
&&
[w^{(1)}_{t_2} , \tilde B^{(1)}]  + w^{(1)}_{t_1t_1}- \alpha [w^{(1)}_{t_1} w^{(1)},\tilde B^{(1)}] + 
\alpha\Big([w^{(1)},\tilde B^{(1)}] w^{(1)}\Big)_{t_1}=0
\end{eqnarray}
If, in addition, 
\begin{eqnarray}\label{M_eq_2}\label{M_eq_2_B3}
M=2,\;\; \tilde B^{(1)}={\mbox{diag}}(1,-1), 
\end{eqnarray}
this equation  yields the celebrated NLS system:
 \begin{eqnarray}\label{NLS}
&&
\frac{1}{\alpha}q_{t_2} -\frac{1}{2} q_{\tau_1\tau_1}  - 4 r q^2=0,
\\\nonumber
&&
\frac{1}{\alpha}r_{t_2} +\frac{1}{2} r_{\tau_1\tau_1} +  4 q r^2=0
\end{eqnarray}
for the off-diagonal elements of $w^{(1)}$: $q=w^{(1)}_{12},\;\; r=w^{(1)}_{21}$. The NLS equation 
$iq_{t_2}+\frac{1}{2} q_{\tau_1\tau_1}+  4 q^2\bar q=0$ corresponds to the reduction $r=\bar q$, $\alpha=i$.

\paragraph{KdV and mKdV.} 
Eq.(\ref{n_wave_ch_11}) with $k=1$, $n=1,2,3$, and eq.(\ref{Burgers-3_ch_11}) with $n=1$ yield
 a complete system of PDEs for $w^{(j)}$, $j=1,2,3,4$:
\begin{eqnarray}\label{KdV_ch}
&&
w^{(1)}_{t_1} +[w^{(2)},\tilde B^{(1)}] +[w^{(1)},\tilde B^{(1)}] w^{(1)} =0,\\\nonumber
&&
w^{(2)}_{t_1} +[w^{(3)},\tilde B^{(1)}] +[w^{(1)},\tilde B^{(1)}] w^{(2)} =0,\\\nonumber
&&
w^{(3)}_{t_1} +[w^{(4)},\tilde B^{(1)}] +[w^{(1)},\tilde B^{(1)}] w^{(3)} =0,\\
\nonumber
&&
w^{(1)}_{t_3} +  [w^{(1)},\tilde B^{(3)}] \Big( w^{(1)} (      
w^{(1)} w^{(1)}+  w^{(2)})+w^{(2)}w^{(1)}+w^{(3)}\Big)+ \\\nonumber
&&
[w^{(2)},\tilde B^{(3)}] \Big(w^{(1)} w^{(1)}+ w^{(2)}\Big)+     
[w^{(3)},\tilde B^{(3)}] w^{(1)}+[w^{(4)},\tilde B^{(3)}] =0 .
\end{eqnarray} 
In the case $\tilde B^{(3)}=-\tilde  B^{(1)}$, this system reduces to the two coupled matrix equations
\beq\label{t3}
\ba{l}
w^{(1)}_{t_1} +[w^{(2)},\tilde B^{(1)}] +[w^{(1)},\tilde B^{(1)}] w^{(1)} =0,    \\
-[\tilde B^{(1)},w^{(1)}_{t_1}]=w^{(2)}_{t_1t_1}+
[\tilde B^{(1)},w^{(1)}_{t_1}(w^{(1)}w^{(1)}+w^{(2)})+w^{(2)}_{t_1}w^{(1)}]
-\left([\tilde B^{(1)},w^{(1)}]w^{(2)} \right)_{t_1}.
\ea
\eeq 
If, in addition, the  choice (\ref{M_eq_2_B3}) is made, the system (\ref{t3}) becomes the mKdV system:
\begin{eqnarray}\label{KdV_mKdV}
&&
q_{t_3}+\frac{1}{4}q_{t_1t_1t_1}+ 6 q_{t_1} qr=0,\\\nonumber
&&
r_{t_3}+\frac{1}{4}r_{t_1t_1t_1}+ 6 r_{t_1} rq=0,
\end{eqnarray}
where $q=w^{(1)}_{12},\;\;r=w^{(1)}_{21}$, reducing to the KdV equation $q_{t_3}+\frac{1}{4}q_{t_1t_1t_1}+ 6 q q_{t_1}=0$ and to the 
mKdV equation $q_{t_3}+\frac{1}{4}q_{t_1t_1t_1}+ 6 q^2 q_{t_1}=0$ if $r=1$ and $r=q$ respectively.

\subsubsection{Lax pairs for $S$-integrable PDEs in (1+1)-dimensions}
\label{Section:Lax_1+1}
As in \S 2, in order to derive the Lax pairs for the above $S$-integrable PDEs in (1+1)-dimensions, it is convenient to      
write the system (\ref{sp_11},\ref{sp_t_m}) in the equivalent form 
\begin{eqnarray}\label{sp_11m}
&&
 w\Psi=\Psi \Lambda,
 \\\label{sp_112_m}
\label{sp_12_m}
&&
\Psi_{t_{mk}}- B^{(mk)} w^m
\Psi=0,\;\;m\in\NN_+ ,
\end{eqnarray}
and consider the dual system
\begin{eqnarray}\label{sp_11_d}
&&
\tilde \Psi w=\Lambda\tilde \Psi,\\
\label{sp_112_d}\label{sp_12_d}
&&
\tilde \Psi_{t_{mk}}+\tilde \Psi  B^{(mk)} w^{(m)}
=0,
\;\;m\in\NN_+.
\end{eqnarray}
Taking into account the block structure of the matrix $w$ given by eq.(\ref{W}) and considering the
 first rows of eqs.(\ref{sp_11_d},\ref{sp_12_d}), we obtain the following spectral chains, for $n\in\NN_+$:
\begin{eqnarray}\label{sp_x_11}
&&
\tilde \Psi^{(1n)}{\cal E} = \tilde \Psi^{(11)}w^{(n)} +  \tilde \Psi^{(1(n+1))},
\\\label{sp_t12_11}
\label{sp_t12_112}
&&
\tilde  \Psi^{(1n)}_{t_{1k}}+ \tilde \Psi^{(11)} \tilde B^{(1k)} w^{(n)} + \tilde \Psi^{(1(n+1))} \tilde B^{(1k)}=0,\\\label{sp_t3_11} 
 && \tilde \Psi^{(1n)}_{t_{2k}}+ \tilde \Psi^{(11)}  \tilde B^{(2k)} (w^{(1)}w^{(n)}+w^{(n+1)}) +
\tilde  \Psi^{(12)} \tilde B^{(2k)} w^{(n)} +\tilde  \Psi^{(1(n+2))}\tilde B^{(2k)}=0,\\ \label{sp_t5_11}
&& 
\tilde \Psi^{(1n)}_{t_{3k}} +\tilde \Psi^{(11)} \tilde B^{(3k)} 
\Big( w^{(1)}( w^{(1)} w^{(n)}+w^{(n+1)})+ w^{(2)}w^{(n)}+w^{(n+2)}\Big)  \\\nonumber
&&
+\tilde \Psi^{(12)} \tilde B^{(3k)} \Big(w^{(1)} w^{(n)}+ w^{(n+1)}\Big)
+ \tilde \Psi^{(13)} \tilde B^{(3k)} w^{(n)}+\tilde \Psi^{(1(n+3))} \tilde B^{(3k)} = 0.
\end{eqnarray} 
where ${\cal E}=\Lambda^{(1)}$. Setting $n=1$ into eqs.(\ref{sp_x_11},\ref{sp_t12_112}) and eliminating $\tilde \Psi^{(12)}$,
 one gets the dual of the Lax pair for the (1+1)-dimensional $N$-wave equation (\ref{S_n_w_11}) 
($\tilde \psi=\tilde \Psi^{(11)}$):
 \begin{eqnarray}
\label{sp2_t12_11} \label{sp2_t12_112}
&&
\tilde\psi_{t_{1k}}+{\cal E} \tilde\psi\tilde B^{(1k)}+ \tilde\psi [\tilde B^{(1k)}, w^{(1)}] =0,\;\;k=1,2
.
\end{eqnarray} 
Eq.(\ref{sp2_t12_11}) with $k=1$, written in terms of (\ref{red_var}), is the first  equation of the dual Lax pair also for eqs.(\ref{S_11}) and (\ref{KdV_ch}).

Setting $k=n=1$ into eq.(\ref{sp_t3_11}) and eliminating $\tilde \Psi^{(12)}$, $\tilde \Psi^{(13)}$ using eq.(\ref{sp_x_11}), 
one gets the second equation of the dual Lax pair for (\ref{S_11}):
 \begin{eqnarray}
\label{sp2_t12_NLS} 
  &&
  \tilde\psi_{t_2}+{\cal E}^2 \tilde\psi\tilde B^{(2)}+{\cal E} \tilde\psi [\tilde B^{(2)}, w^{(1)}] + \tilde\psi s=0,\;\;\\\nonumber
&&
 s= [\tilde B^{(2)},w^{(2)}]  +[\tilde B^{(2)} , w^{(1)}] w^{(1)}.
\end{eqnarray} 

The second equation of the dual Lax pair for the eq.(\ref{KdV_ch}) results from the eq.(\ref{sp_t5_11}), $k=n=1$ after eliminating 
$\tilde \Psi^{(1j)}$, $j=2,3,4$ using eq.(\ref{sp_x_11}). In view of conditions  (\ref{M_eq_2_B3}) the complete dual  
spectral system    reads:
\begin{eqnarray}\label{sp_KdV}
&&
\tilde\psi_{t_1} +{\cal E}\tilde \psi
\left[
\begin{array}{cc}
1&0\cr
0&-1
\end{array}\right]+\tilde \psi
\left[
\begin{array}{cc}
0&2q\cr
-2r&0
\end{array}
\right]=0,\\\nonumber
&&
\tilde\psi_{t_3} -{\cal E}^3\tilde \psi
\left[
\begin{array}{cc}
1&0\cr
0&-1
\end{array}\right]-{\cal E}^2\tilde \psi
\left[
\begin{array}{cc}
0&2q\cr
-2r&0
\end{array}
\right]+\\\nonumber
&&
{\cal E}\tilde \psi\left[
\begin{array}{cc}
-2 qr&-q_{\tau_1}\cr
-r_{\tau_1}&2 qr
\end{array}
\right]-\tilde \psi\left[
\begin{array}{cc}
rq_{\tau_1}- qr_{\tau_1}&
\frac{1}{2}q_{\tau_1\tau_1} + 4 q^2 r\cr
-\frac{1}{2}r_{\tau_1\tau_1} - 4 r^2 q& 
qr_{\tau_1}-rq_{\tau_1}
\end{array}
\right]=0
\end{eqnarray}

The duals of the equations (\ref{sp2_t12_11}-\ref{sp_KdV}) read:
\begin{eqnarray}
\label{ch_sp2_t12_11}
 \label{ch_sp2_t12_112}
&&
 \psi_{t_{1k}}- \tilde B^{(1k)}\psi{\cal E}-[\tilde B^{(1k)} ,w^{(1)}]\psi  =0,\;\;k=1,2,
\\
\label{ch_sp2_t12_NLS} 
  &&
  \psi_{t_2}- \tilde B^{(2)} \psi {\cal E}^2-[\tilde B^{(2)}, w^{(1)}]  \psi{\cal E} -s\psi =0,
\end{eqnarray} 
\begin{eqnarray}\label{ch_sp2_t12_KdV}
&&
\psi_{t_1} +
\left[
\begin{array}{cc}
1&0\cr
0&-1
\end{array}\right] \psi{\cal E}-
\left[
\begin{array}{cc}
0&2q\cr
-2r&0
\end{array}
\right]\psi=0,\\\nonumber
&&
\psi_{t_3}-
\left[
\begin{array}{cc}
1&0\cr
0&-1
\end{array}\right]\psi{\cal E}^3+
\left[
\begin{array}{cc}
0&2q\cr
-2r&0
\end{array}
\right]\psi{\cal E}^2+\\\nonumber
&&
 \left[
\begin{array}{cc}
-2 qr&-q_{\tau_1}\cr
-r_{\tau_1}&2 qr
\end{array}
\right]\psi{\cal E}+
\left[
\begin{array}{cc}
rq_{\tau_1}- qr_{\tau_1}&
\frac{1}{2}q_{\tau_1\tau_1} + 4 q^2 r\cr
-\frac{1}{2}r_{\tau_1\tau_1} - 4 r^2 q& 
qr_{\tau_1}-rq_{\tau_1}
\end{array}
\right]
 \psi=0
\end{eqnarray}
Eqs. (\ref{ch_sp2_t12_11}) are the Lax pair of the $N$-wave eq.(\ref{S_n_w_11}), 
eq. (\ref{ch_sp2_t12_11}) with $k=1$ and eq.(\ref{ch_sp2_t12_NLS}) are the Lax pair of 
eq.(\ref{off_11}), reducing to the NLS system if (\ref{M_eq_2}) holds, and eqs.(\ref{ch_sp2_t12_KdV}) are Lax pair of  the system (\ref{KdV_mKdV}) reducing to either KdV or mKdV.

\subsubsection{From the Lax pairs of (1+1)-dimensional $S$-integrable PDEs to the Ch-integrable eqs.(\ref{wS1})}
We show that the eqs.(\ref{wS1})   may be derived from the spectral problems obtained in 
$\S$\ref{Section:Lax_1+1}. 
Let
 \begin{eqnarray}\label{ch_UV}
 \psi(\Lambda;x) =  \chi (\Lambda;x)e^{\sum\limits_{i,k\ge 1}(-1)^i 
 \tilde B^{(ik)} t_{ik}\Lambda^i }.
\end{eqnarray} 
Then equation (\ref{ch_sp2_t12_11}-\ref{ch_sp2_t12_KdV}) yield:
\begin{eqnarray}
\label{chi_ch_sp2_t12_11}
 \label{chi_ch_sp2_t12_112}
&&
 \chi_{t_{1k}}+\Lambda [\tilde B^{(1k)},\chi]-
 [\tilde B^{(1k)} ,w^{(1)}]\chi  =0
,\;\;k=1,2,\\
\label{chi_ch_sp2_t12_NLS} 
  &&
  \chi_{t_2}-
  \Lambda^2 [\tilde B^{(2)}, \chi]-
  [ w^{(1)},\tilde B^{(2)}] \Lambda \chi -s\chi =0,\;\;\\\nonumber
&&
 s= [\tilde B^{(2)},w^{(2)}]  +
 [\tilde B^{(2)} , w^{(1)}] w^{(1)}.
\end{eqnarray} 
\begin{eqnarray}&&
\chi_{t_1} +
\left[\left[
\begin{array}{cc}
1&0\cr
0&-1
\end{array}\right] ,\chi\right]\Lambda-
\left[
\begin{array}{cc}
0&2q\cr
-2r&0
\end{array}
\right]\chi=0,\\\label{ch_sp2_t12_KdV_chi}
&&
\chi_{t_3} - \left[
\left[
\begin{array}{cc}
1&0\cr
0&-1
\end{array}\right],\chi\right]\Lambda^3+
\left[
\begin{array}{cc}
0&2q\cr
-2r&0
\end{array}
\right]\chi\Lambda^2+\\\nonumber
&&
\left[
\begin{array}{cc}
-2 qr&-q_{\tau_1}\cr
-r_{\tau_1}&2 qr
\end{array}
\right]\chi\Lambda+\left[
\begin{array}{cc}
rq_{\tau_1}- qr_{\tau_1}&
\frac{1}{2}q_{\tau_1\tau_1} + 4 q^2 r\cr
-\frac{1}{2}r_{\tau_1\tau_1} - 4 r^2 q& 
qr_{\tau_1}-rq_{\tau_1}
\end{array}
\right] \chi
\end{eqnarray}
It is now easy to verify, as in \S 2, that the coefficients of the $\Lambda$ - large expansion of 
the eigenfunction $\chi$ satisfy the infinite chains (\ref{n_wave_ch_11}), (\ref{n_wave_ch_111}), 
(\ref{Burgers_ch_11}), (\ref{Burgers-3_ch_11}):
\begin{eqnarray}\label{inf}
\chi(\Lambda;y) \to I -\sum_{j\ge 1}(-1)^j w^{(j)}\Lambda^{-j},
\end{eqnarray}
clarifying their spectral meaning. In addition, recompiling such coefficients into the block Frobenius matrix, we 
reconstruct the matrix equations (\ref{wS1}) from the Lax pairs of (1+1)-dimensional $S$-integrable PDEs.  

\subsection{Derivation of the SDYM equation and of its multi-dimensional generalizations}

Now we take, in equations (\ref{H_sp}),(\ref{H_flows}), $B^{(nk)}=0$ (or $\sigma^{(kj)}=0$), $m=1$, 
$\rho^{(1jk)}(\Lambda)=\Lambda^j$, $t_{11}=t$, $t_{12}=\tau$, $x_{j1}=x_j$, $x_{j2}=y_j$, obtaining the system
\begin{eqnarray}\label{gg_sp_ch}
&&
w\Psi= \Psi \Lambda,           \\\nonumber
&&
\Psi_{t} +\sum_{j=1}^N\Psi_{x_j}\Lambda^j=0,\\\nonumber
&&
\Psi_{\tau} +\sum_{j=1}^N\Psi_{y_j}\Lambda^j =0,
\end{eqnarray}
 whose compatibility condition yields
\begin{eqnarray}\label{g_char_w}
&&
w_{t} + \sum_{j=1}^N w_{x_j} w^j =0,\\\nonumber
&&
w_{\tau} +\sum_{j=1}^N w_{y_j} w^j =0.
\end{eqnarray}
We proceed as in the previous sections but, before considering the derivation in the general case, quite complicated, we 
illustrate the simplest two examples.

\subsubsection{$N=1$: the $GL(M,\CC)-SDYM$ equation}
The compatibility condition of the system (\ref{gg_sp_ch}) yields
\begin{eqnarray}\label{char_w}
w_{t} + w_{x_1} w =0,\;\;\;w_{\tau} + w_{y_1} w =0 .
\end{eqnarray}
Let $w$ and $\Psi$ be given by eqs.(\ref{W}) and (\ref{Psi13}) respectively.
The first rows of the matrix equations (\ref{char_w}) generate the chains, for $n\in\NN_+$ \cite{Z4}: 
\begin{eqnarray}\label{P_ch}
&&
w^{(n)}_{t} + w^{(1)}_{x_1} w^{(n)}+ w^{(n+1)}_{x_1}=0,\\\nonumber
&&
w^{(n)}_{\tau} + w^{(1)}_{y_1} w^{(n)}+ w^{(n+1)}_{y_1}=0.
\end{eqnarray}
Setting $n=1$ and eliminating $w^{(2)}$, we derive the well-known $GL(M,\CC)-SDYM$ equation:
\begin{eqnarray}\label{SDYM}
w^{(1)}_{t y_1}-w^{(1)}_{\tau x_1} + [w^{(1)}_{x_1}, w^{(1)}_{y_1}]=0.
\end{eqnarray}

To derive the Lax pair of (\ref{SDYM}), we write first the dual of system (\ref{gg_sp_ch}) in the convenient form:
\begin{eqnarray}\label{sp_ch_d}
&&
\tilde \Psi w=\Lambda \tilde\Psi ,\\\nonumber
&&
\tilde \Psi_{t} +\tilde \Psi_{x_1}w =0,~~~~~~~\tilde \Psi_{\tau} +\tilde \Psi_{y_1}w =0.
\end{eqnarray}
Using again (\ref{W}) and (\ref{Psi13}),  
the first rows of equations (\ref{sp_ch_d}) appear in the form:
\begin{eqnarray}\label{sp_ch_d_ch}
&&
\tilde \Psi^{(11)}
w^{(n)}+\tilde \Psi^{(1(n+1))}={\cal{E}} \tilde\Psi^{(1n)} ,\\\label{sp_sd}\nonumber
&&
\tilde \Psi^{(1n)}_{t} + \tilde \Psi^{(11)}_{x_1} w^{(n)}+ \tilde \Psi^{(1(n+1))}_{x_1}=0,\\\nonumber
&&
\tilde \Psi^{(1n)}_{\tau} + \tilde \Psi^{(11)}_{y_1} w^{(n)}+ \tilde \Psi^{(1(n+1))}_{y_1}=0,
\end{eqnarray}
where ${\cal E}=\Lambda^{(1)}$. Setting $n=1$ in eqs.(\ref{sp_ch_d_ch}) and eliminating $\tilde \Psi^{(12)}$, 
one  obtains the dual of the Lax pair of (\ref{SDYM}) for the spectral function  $\tilde\psi=\tilde \Psi^{(11)}$:
\begin{eqnarray}\label{Sd_dual_Lax}
&&
\tilde\psi_{t} + ({\cal E} \tilde\psi)_{x_1} = \tilde\psi w^{(1)}_{x_1},\\\nonumber
&&
\tilde\psi_{\tau} + ({\cal E} \tilde\psi)_{y_1} = \tilde\psi w^{(1)}_{y_1}.
\end{eqnarray}
Then the Lax pair of (\ref{SDYM}) reads:
\begin{eqnarray}\label{Lax_SD}
&&
\psi_{t} + \psi_{x_1}{\cal E} + w^{(1)}_{x_1}\psi =0,\;\;
\\\nonumber
&&
\psi_{\tau} + \psi_{y_1}{\cal E} + w^{(1)}_{y_1}\psi =0.
\end{eqnarray}

Vice-versa, it is easy to verify that the coefficients of the ${\cal{E}}$ large expansion of 
the eigenfunction $\psi$ in (\ref{Lax_SD}) 
are the elements of the chain (\ref{P_ch}):
\begin{eqnarray}\label{ch_inf}
\psi({\cal{E}};\vec{x}) \to I -\sum_{j\ge 1} w^{(j)}{\cal{E}}^{-j},
\end{eqnarray}
obtaining the spectral meaning of such chains. As a consequence,  
one reconstructs eqs.(\ref{char_w}) from the Lax pair (\ref{Lax_SD}) of the SDYM equation.

\subsubsection{$N=2$: a generalization of the $SDYM$ equation in $6$ dimensions}

The compatibility condition of the system (\ref{gg_sp_ch}) yields now
\begin{eqnarray}\label{2_w}
w_{t} + w_{x_1} w +w_{x_2} w^2=0,\;\;\;w_{\tau} + w_{y_1} w+w_{y_2} w^2 =0 .
\end{eqnarray}

The corresponding chains read:
\begin{eqnarray}\label{2_ch}
&&
w^{(n)}_{t} + w^{(1)}_{x_1} w^{(n)}+ w^{(n+1)}_{x_1}+ 
w^{(1)}_{x_2}(w^{(1)}w^{(n)}+w^{(n+1)})+w^{(2)}_{x_2}w^{(n)}+ w^{(n+2)}_{x_2}=0,\\\nonumber
&&
w^{(n)}_{\tau} + w^{(1)}_{y_1} w^{(n)}+ w^{(n+1)}_{y_1}+ w^{(1)}_{y_2}(w^{(1)}w^{(n)}+w^{(n+1)})+
w^{(2)}_{y_2}w^{(n)}+ w^{(n+2)}_{y_2}=0.
\end{eqnarray}
Setting $n=1$ and $n=2$ in (\ref{2_ch}) and eliminating the fields $w^{(3)},w^{(4)}$, one obtains the following 
integrable system of two nonlinear PDEs in $6$ dimensions for the fields $w^{(1)},w^{(2)}$:
\beq\label{2_SDYM}
\ba{l}
w^{(1)}_{x_2\tau} -w^{(1)}_{y_2t}+w^{(2)}_{x_2y_1}-w^{(2)}_{x_1y_2}+w^{(1)}_{x_2y_1}w^{(1)}-w^{(1)}_{x_1y_2}w^{(1)}+
w^{(1)}_{y_1}w^{(1)}_{x_2}-w^{(1)}_{x_1}w^{(1)}_{y_2}+[w^{(2)}_{y_2},w^{(1)}_{x_2}]  \\
 + [w^{(1)}_{y_2},w^{(2)}_{x_2}]+
w^{(1)}_{y_2}\left({w^{(1)}}^2\right)_{x_2}-w^{(1)}_{x_2}\left({w^{(1)}}^2\right)_{y_2}=0, \\
~~  \\
 w^{(1)}_{x_1\tau} -w^{(1)}_{y_1t}+ [w^{(1)}_{y_1},w^{(1)}_{x_1}] + w^{(2)}_{x_2\tau}-w^{(2)}_{y_2 t}+
[w^{(2)}_{y_2},w^{(1)}_{x_1}]+[w^{(1)}_{y_1},w^{(2)}_{x_2}]+[w^{(1)}_{x_2},w^{(1)}_{y_2}]{w^{(1)}}^2+ \\
\left(w^{(1)}_{x_1y_2}w^{(1)}-w^{(1)}_{x_2y_1}w^{(1)}+
w^{(2)}_{x_1y_2}-w^{(2)}_{x_2y_1}\right)w^{(1)}+ w^{(1)}_{x_2}\left(w^{(1)}_{\tau}-
w^{(1)}w^{(1)}_{y_1}+[w^{(2)}_{y_2},w^{(1)}]\right)- \\
w^{(1)}_{y_2}\left(w^{(1)}_{t}-w^{(1)}w^{(1)}_{x_1}+[w^{(2)}_{x_2},w^{(1)}]\right)+[w^{(2)}_{y_2},w^{(2)}_{x_2}]=0,
\ea
\eeq
reducing to (\ref{SDYM}) for $w^{(1)}$ if the fields do not depend on ${x_2},y_2$.

To derive the Lax pair of (\ref{2_SDYM}), we consider again the dual of system (\ref{gg_sp_ch}):
\begin{eqnarray}\label{sp_ch_d_2}
&&
\tilde \Psi w=\Lambda \tilde\Psi ,\\\nonumber
&&
\tilde \Psi_{t} +\tilde \Psi_{x_1}w+\tilde \Psi_{x_2}w^2 =0,~~~~~~~\tilde \Psi_{\tau} +\tilde \Psi_{y_1}w+\tilde \Psi_{y_2}w^2 =0.
\end{eqnarray}
Using (\ref{W}) and (\ref{Psi13}),  
the first rows of equations (\ref{sp_ch_d_2}) appear in the form:
\begin{eqnarray}\label{sp_ch_d_ch_2}
&&
\tilde \Psi^{(11)}w^{(n)}+\tilde \Psi^{(1(n+1))}={\cal{E}} \tilde\Psi^{(1n)} ,\\\label{sp_sd_2_t}
&&
\tilde \Psi^{(1n)}_{t} + \tilde \Psi^{(11)}_{x_1} w^{(n)}+ \tilde \Psi^{(1(n+1))}_{x_1}+\tilde \Psi^{(11)}_{x_2}(w^{(1)}w^{(n)}+w^{(n+1)})+
\tilde \Psi^{(12)}_{x_2}w^{(n)}+\tilde \Psi^{(1(n+2))}_{x_2} =0,\\\label{sp_sd_2_tau}
&&
\tilde \Psi^{(1n)}_{\tau} + \tilde \Psi^{(11)}_{y_1} w^{(n)}+ \tilde \Psi^{(1(n+1))}_{y_1}+\tilde \Psi^{(11)}_{y_2}(w^{(1)}w^{(n)}+w^{(n+1)})+
\tilde \Psi^{(12)}_{y_2}w^{(n)}+\tilde \Psi^{(1(n+2))}_{y_2} =0.
\end{eqnarray}
Setting $n=1$ in equations (\ref{sp_sd_2_t},\ref{sp_sd_2_tau}), and eliminating $\tilde \Psi^{(12)},\tilde \Psi^{(13)}$    
using  eqs.(\ref{sp_ch_d_ch_2}) for $n=1,2$,  
one  obtains the dual of the Lax pair of (\ref{2_SDYM}) for the spectral function  $\tilde\psi=\tilde \Psi^{(11)}$:
\begin{eqnarray}\label{Sd_dual_Lax2}
&&
\tilde \psi_{t} +\sum_{j=1}^2({\cal{E}}^j \tilde \psi)_{x_j}=\tilde \psi\left(w^{(1)}_{x_1}+w^{(2)}_{x_2}+w^{(1)}_{x_2}w^{(1)}\right)+
 {\cal{E}} \tilde \psi w^{(1)}_{x_2} ,\\\nonumber
&&
\tilde \psi_{t} +\sum_{j=1}^2({\cal{E}}^j \tilde \psi)_{y_j}=\tilde \psi\left(w^{(1)}_{y_1}+w^{(2)}_{y_2}+w^{(1)}_{y_2}w^{(1)}\right)+
 {\cal{E}} \tilde \psi w^{(1)}_{y_2} .
\end{eqnarray}
Then the Lax pair of (\ref{2_SDYM}) reads:
\begin{eqnarray}\label{Lax_SD_2}
&&
\psi_{t} + \psi_{x_1}{\cal E}+ \psi_{x_2}{\cal E}^2+\left(w^{(2)}_{x_2}+w^{(1)}_{x_2}w^{(1)}+w^{(1)}_{x_1}\right)\psi+w^{(1)}_{x_2}\psi{\cal E} =0,\;\;
\\\nonumber
&&
\psi_{\tau} +\psi_{y_1}{\cal E}+ \psi_{y_2}{\cal E}^2+\left(w^{(2)}_{y_2}+w^{(1)}_{y_2}w^{(1)}+w^{(1)}_{y_1}\right)\psi+w^{(1)}_{y_2}\psi{\cal E}  =0.
\end{eqnarray}

As before, it is easy to verify that equation (\ref{ch_inf}) holds, namely that 
the coefficients of the ${\cal{E}}$ large expansion of $\psi$ in (\ref{Lax_SD_2})
are the elements of the chain (\ref{2_ch}).
Therefore one reconstructs equations (\ref{2_w}) from the Lax pair (\ref{2_SDYM}) of the six dimensional generalization 
(\ref{2_SDYM}) of the SDYM equation.

\subsubsection{Multidimensional generalization of the $SDYM$ equation}

Motivated by the above formulae for the simplest cases $N=1,2$, here we discuss the general $N$ situation. 
If $w$ is the block Frobenius matrix (\ref{W}), the power $w^j$ exhibits the following structure
\begin{eqnarray}\label{WW^j}
&&
w^j=\left[\begin{array}{cccc}
\tilde w^{(j;11)} & \tilde w^{(j;12)} & \tilde w^{(j;13)} &\cdots\cr
\cdots&\cdots&\cdots&\cdots\cr
\tilde w^{(j;j1)} & \tilde w^{(j;j2)} & \tilde w^{(j;j3)} &\cdots\cr
I_M&0_M&0_M&\cdots\cr
0_M&I_M&0_M&\cdots\cr
\vdots&\vdots&\vdots&\ddots
\end{array}\right],
\end{eqnarray}
where the matrix blocks $\tilde w$ are defined by the equations
\begin{eqnarray}
\label{def_w_tilde1}
&&
\tilde w^{(j;1n)}=\sum\limits_{i=1}^{j-1}w^{(i)} \tilde w^{(j-i;1n)} + w^{(j+n-1)}, ~~ n,j\ge 1, ~~ (\tilde w^{(1;1n)}=w^{(n)}), 
\\\label{def_w_tilde2}
&&
\tilde w^{(j;kn)}=\tilde w^{(j-k+1;1n)}, ~~~~~~~~~~~~~~~~~~~~~~ 2\le k\le j, ~~~(\tilde w^{(j;jn)}=w^{(n)})
\end{eqnarray} 
in terms of the basic blocks $w^{(j)},~j\ge 1$. The first few examples read:
\begin{eqnarray}
\tilde w^{(2;1n)}&=&w^{(1)}w^{(n)} + w^{(n+1)},\\\nonumber
\tilde w^{(3;1n)}&=&(w^{(1)})^2w^{(n)} + w^{(1)} w^{(n+1)}+ 
 w^{(2)} w^{(n)}+ w^{(n+2)},\\\nonumber
\tilde w^{(4;1n)}&=&w^{(1)}\Big((w^{(1)})^2w^{(n)} + w^{(1)} w^{(n+1)}+ 
w^{(2)} w^{(n)}+ w^{(n+2)}\Big)+\\\nonumber
&& w^{(2)} \Big(w^{(1)}w^{(n)} + w^{(n+1)}\Big) + w^{(3)} w^{(n)}
+ w^{(n+3)}.
\end{eqnarray}

Furthermore, evaluating the $(1n)$-block of $w^{j+1}$, written as $(w^{j}w)$, we obtain the additional 
formula
\beq
\tilde w^{(j-1;1(n+1))}=\tilde w^{(j;1n)}-\tilde w^{(j-1;11)}w^{(n)},~~j\ge 1,\;\;n>1,
\eeq
implying
\beq
\label{w_tilde}
\tilde w^{(j;1n)}=\tilde w^{(j+n-1;11)}-\sum\limits_{l=1}^{n-1}\tilde w^{(j+l-1;11)}w^{(n-l)},~~j,n\ge 1.
\eeq
Eq.(\ref{w_tilde}) reduces, for $j=1$, to the following equation:
\beq
\label{new}
\tilde w^{(j;11)}=\sum\limits_{i=1}^{j-1}\tilde w^{(j-i;11)}w^{(i)}+w^{(j)},~~j\ge 1
\eeq
useful later on (compare it with equation (\ref{def_w_tilde1}) for $n=1$).

Using equation (\ref{W}), the system (\ref{g_char_w}) generates the discrete chains
\begin{eqnarray}\label{g_chain}
&&
w^{(n)}_{t} + \sum_{j=1}^N\Big(\sum_{i=1}^j w^{(i)}_{x_j} \tilde w^{(j-i+1;1n)} +
 w^{(j+n)}_{x_j} \Big) =0,\\\nonumber
&&
w^{(n)}_{\tau} + \sum_{j=1}^N\Big(\sum_{i=1}^j w^{(i)}_{y_j} \tilde w^{(j-i+1+;in)} + w^{(j+n)}_{y_j} \Big) =0.
\end{eqnarray}
Setting $n=1,\dots,N$ in equations (\ref{g_chain}), one obtains a determined system of $2N$ equations for the fields  
$w^{(i)}$, $i=1,\dots,2N$. As in the previous two illustrative examples for $N=1,2$, it is possible to eliminate the $N$ fields 
$w^{(i)}$, $i=N+1,\dots,2N$, obtaining a system on $N$ equations in ($2N+2$) dimensions for the remaining 
fields $w^{(i)}$, $i=1,\dots,N$. Such system, which provides the natural multidimensional generalization 
of the SDYM equation (\ref{SDYM}), is conveniently written as follows
\beq
\label{SDYM_gen}
\ba{l}
p^{(N,0)}_{\tau} - q^{(N,0)}_{t}+[q^{(0)},p^{(0)}]=0,                                               \\
p^{(N,n)}_{\tau} - q^{(N,n)}_{t}+\sum\limits_{j=1}^n\left(p^{(N,n-j)}_{y_j}-q^{(N,n-j)}_{x_j}\right)
 +\sum\limits_{j=0}^n\left[q^{(N,j)},p^{(N,n-j)}\right]=0,  \;\;              1\le n\le N-1,
\ea
\eeq
where the fields $p^{(N,j)},q^{(N,j)}$ are suitable combinations of the $N$ fields $w^{(n)},~n=1,\dots,N$:
\beq\label{def_pq}
\ba{l}
p^{(N,j)}=\sum\limits_{s=j+1}^N\left(\sum\limits_{l=1}^{s-j-1}w^{(l)}_{x_s}\tilde w^{(s-j-l;11)}+
w^{(s-j)}_{x_s}\right),  \\
q^{(N,j)}=\sum\limits_{s=j+1}^N\left(\sum\limits_{l=1}^{s-j-1}w^{(l)}_{y_s}\tilde w^{(s-j-l;11)}+
w^{(s-j)}_{y_s}\right);
\ea
\eeq
the first few read as follows:
\beq
\ba{l}
p^{(N,N-1)}=w^{(1)}_{x_N},~~~~~~~~~~~~~~~~~~~~~~~~~~~~q^{(N,N-1)}=w^{(1)}_{y_N},        \\
p^{(N,N-2)}=w^{(1)}_{x_{N-1}}+w^{(2)}_{x_N}+w^{(1)}_{x_N}w^{(1)},~~~
q^{(N,N-2)}=w^{(1)}_{y_{N-1}}+w^{(2)}_{y_N}+w^{(1)}_{y_N}w^{(1)},  \\
p^{(N,N-3)}=w^{(1)}_{x_{N-2}}+w^{(2)}_{x_{N-1}}+w^{(3)}_{x_N}+\left(w^{(1)}_{x_{N-1}}+w^{(2)}_{x_N}\right)w^{(1)}+
w^{(1)}_{x_N}\left({w^{(1)}}^2+w^{(2)}\right),                     \\
q^{(N,N-3)}=w^{(1)}_{y_{N-2}}+w^{(2)}_{y_{N-1}}+w^{(3)}_{y_N}+\left(w^{(1)}_{y_{N-1}}+w^{(2)}_{y_N}\right)w^{(1)}+
w^{(1)}_{y_N}\left({w^{(1)}}^2+w^{(2)}\right).
\ea
\eeq

To show it, it is more convenient to go through the Lax pair derivation.  
 
\paragraph{Lax pair.}
The system dual of (\ref{gg_sp_ch}) reads
\begin{eqnarray}\label{g_sp_ch_d}
&&
\tilde \Psi w=\Lambda \tilde\Psi,    \\\nonumber
&&
\tilde \Psi_{t} +\sum_{j=1}^N
\tilde \Psi_{x_j}w^j =0,~~~\tilde \Psi_{\tau} +\sum_{j=1}^N\tilde \Psi_{y_j} w^j =0.
\end{eqnarray}
and it is conveniently rewritten in the equivalent form
\begin{eqnarray}\label{gg_sp_ch_d}
&&  
\tilde \Psi w=\Lambda \tilde\Psi ,                                    \\\label{aa1}
&&
\tilde \Psi_{t} +\sum_{j=1}^N(\Lambda^j \tilde \Psi)_{x_j}-\sum_{j=1}^N\tilde \Psi (w^j)_{x_j}=
\tilde \Psi_{t} +\sum_{j=1}^N(\Lambda^j \tilde \Psi)_{x_j}
-\sum_{s=0}^{N-1}\Lambda^s\tilde\Psi\sum_{j=s+1}^Nw_{x_j}w^{j-s-1}=0,  \\\label{aa2}
&&
\tilde \Psi_{\tau} +\sum_{j=1}^N(\Lambda^j\tilde \Psi)_{y_j}-\sum_{j=1}^N\tilde \Psi (w^j)_{y_j}=
\tilde \Psi_{t} +\sum_{j=1}^N(\Lambda^j \tilde \Psi)_{y_j}
-\sum_{s=0}^{N-1}\Lambda^s\tilde\Psi\sum_{j=s+1}^Nw_{y_j}w^{j-s-1}=0.
\end{eqnarray}
As before, the block ($1,1$) of the matrix equations (\ref{aa1}),(\ref{aa2}) leads to the dual of the 
Lax pair (${\cal{E}}=\Lambda^{(1)}$) of the multidimensional SDYM equations:
\begin{eqnarray}\label{sp_sd_d1}
&&
\tilde \psi_{t} +\sum_{j=1}^N
({\cal{E}}^j \tilde \psi)_{x_j}=\sum_{j=0}^{N-1} {\cal{E}}^j \tilde \psi p^{(N,j)},  \\\label{sp_sd_d2}
&&
\tilde \psi_{\tau} +\sum_{j=1}^N
({\cal{E}}^j \tilde \psi)_{y_j}=\sum_{j=0}^{N-1} {\cal{E}}^j \tilde\psi q^{(N,j)},
\end{eqnarray}
where $p^{(N,j)}$ and $q^{(N,j)}$ are defined in terms of $w^{(i)}$ and their derivatives in (\ref{def_pq}).

Then one derives the corresponding Lax pair
\begin{eqnarray}\label{sp_sd1}
&&
 \psi_{t} +\sum_{j=1}^N\psi_{x_j}{\cal{E}}^j+\sum_{j=0}^{N-1} p^{(N,j)} \psi {\cal{E}}^j =0 ,\\\label{sp_sd2}
&&
\psi_{\tau} +\sum_{j=1}^N \psi_{y_j}{\cal{E}}^j+\sum_{j=0}^{N-1} q^{(N,j)} \psi{\cal{E}}^j =0 ,
\end{eqnarray}
together with its compatibility condition, the following determined system of $2N$ equation in 
($2N+2$) variables for the fields $p^{(N,j)},~q^{(N,j)},~j=0,\dots ,N-1$:    
\begin{eqnarray}
\label{p-q-1}
&&
p^{(N,0)}_{\tau} - q^{(N,0)}_{t}+[q^{(0)},p^{(0)}]=0, 
\\\label{p-q-2}
&&
p^{(N,n)}_{\tau} - q^{(N,n)}_{t}+\sum\limits_{j=1}^n\left(p^{(N,n-j)}_{y_j}-q^{(N,n-j)}_{x_j}\right)
 +\sum\limits_{j=0}^n\left[q^{(N,j)},p^{(N,n-j)}\right]=0,  \; 1\le n\le N-1,
\\\label{p-q-3} 
&&
\sum\limits_{j=n-N+1}^N\left(p^{(N,n-j)}_{y_j}-q^{(N,n-j)}_{x_j}\right)+ 
\sum\limits_{j=n-N+1}^{N-1}\left[q^{(N,j)},p^{(N,n-j)}\right]=0,  \;\;        N \le n \le 2N-2,
\\\label{p-q-4} 
&&
p^{(N,N-1)}_{y_{N}} -q^{(N,N-1)}_{x_N}=0.
\end{eqnarray} 

We remark that only the first $N$ equations (\ref{p-q-1}),(\ref{p-q-2}) involve derivatives with respect to the time 
variables $t,\tau$; the remaining $N$ equations (\ref{p-q-3}),(\ref{p-q-4}), providing a set of relations among the 
$2N$ fields $p^{(N,j)},q^{(N,j)}$, are remarkably parametrized by equations (\ref{def_pq}) in terms of the 
$N$ fields $w^{(j)},~j=1,\dots,N$. Therefore one is left with equations (\ref{SDYM_gen}),(\ref{def_pq}). 

We also remark that the generalization (\ref{p-q-1}-\ref{p-q-4}) of the SDYM equation is known in the literature 
\cite{ZS2,Zakharov2}, to be 
generated by the Lax pair
\begin{eqnarray}\label{LP_old}
&&
 \psi_{t} +\sum_{j=1}^N{\lambda}^j\psi_{x_j}+\sum_{j=0}^{N-1}{\lambda}^j p^{(N,j)} \psi =0 ,\\\label{sp_sd22}
&&
\psi_{\tau} +\sum_{j=1}^N{\lambda}^j \psi_{y_j}+\sum_{j=0}^{N-1}{\lambda}^j q^{(N,j)} \psi =0 ,
\end{eqnarray}
differing from (\ref{sp_sd1},\ref{sp_sd2}) by the fact that here $\lambda$ is just a {\bf constant and scalar spectral parameter}.

Therefore the remarkable derivation of (\ref{p-q-1}-\ref{p-q-4}) from the matrix equations (\ref{g_char_w}) and its integration scheme  
has allowed one to uncover the following two important properties of the system (\ref{p-q-1}-\ref{p-q-4}). 
\begin{itemize}
\item Half of the equations of the system (\ref{p-q-1}-\ref{p-q-4}) (the non evolutionary part) 
can be parametrized in terms of the blocks $w^{(j)},~j=1,..,N$ of the Frobenius matrix $w$, reducing by half the 
number of equations. 
\item Equations (\ref{p-q-1}-\ref{p-q-4}) turn out to be associated with the novel Lax pair (\ref{sp_sd1},\ref{sp_sd2}), 
in which the diagonal matrix ${\cal{E}}$ satisfies, from (\ref{lambda_flows}), the integrable quasi-linear equations
\beq\label{E_flows}
\ba{l}
{\cal E}_{t} + \sum\limits_{j=1}^N {\cal E}_{x_j}{\cal E}^j =0, ~~~~~
{\cal E}_{\tau} + \sum\limits_{j=1}^N {\cal E}_{y_j}{\cal E}^j=0.
\ea
\eeq
Therefore, as it was already observed in \cite{Z4} in the case of the SDYM equation (\ref{SDYM}), the integration scheme 
associated with such a novel Lax pair makes clear the existence of a rich solution space exhibiting interesting phenomena of 
multidimensional wave breaking. A detailed study of these solutions is postponed to a subsequent paper, together with the comparison 
with the finite gap solutions of the SDYM equation constructed in \cite{Korotkin}, and associated with a Riemann surface  
with branch points satisfying equations (\ref{E_flows}) for $N=1$. 
\end{itemize}
 
\paragraph{From the Lax pair of the multidimensional SDYM to the integrable chains (\ref{g_chain})}

As for the particular cases $N=1,2$, in this section we show that the ${\cal E}$-large limit of the Lax pair 
(\ref{sp_sd1},\ref{sp_sd2}) yields the expansion (\ref{ch_inf}) for the eigenfunction $\psi$. Therefore  
the coefficients of the ${\cal E}$-large expansion of the spectral 
function associated with the $S$-integrable multidimensional generalization of the SDYM equations are solutions 
of the nonlinear chains (\ref{g_chain}), providing the spectral meaning to such nonlinear chains. In addition, 
recompiling the matrices $w^{(j)}$ into the block Frobenius matrix $w$, via (\ref{W}), one 
establishes a remarkable relation between the Lax pair of the multidimensional SDYM and the basic matrix equations 
(\ref{g_char_w}), solvable by the method of characteristics.

To show the validity of the expansion (\ref{ch_inf}), we substitute it into the Lax pair (\ref{sp_sd1},\ref{sp_sd2}), obtaining 
the following pairs of equations
\beq\label{coeff+}
\ba{l}
p^{(N,i)}=\sum\limits_{j=i+1}^{N-1}\left(w^{(j-i)}_{x_j}+p^{(N,j)}w^{(j-i)}\right)+w^{(N-i)}_{x_N},~0\le i \le N-1, \\
q^{(N,i)}=\sum\limits_{j=i+1}^{N-1}\left(w^{(j-i)}_{y_j}+q^{(N,j)}w^{(j-i)}\right)+w^{(N-i)}_{y_N},~0\le i \le N-1,
\ea
\eeq
\beq\label{coeff-}
\ba{l}
w^{(i)}_t+\sum\limits_{j=1}^{N-1}\left(w^{(j+i)}_{x_j}+p^{(N,j)}w^{(j+i)}\right)+w^{(N+i)}_{x_N}+p^{(N;0)} w^{(i)}=0,~~i\ge 1 ,          \\
w^{(i)}_{\tau}+\sum\limits_{j=1}^{N-1}\left(w^{(j+i)}_{y_j}+q^{(N,j)}w^{(j+i)}\right)+w^{(N+i)}_{y_N}+q^{(N;0)} w^{(i)}=0,~~i\ge 1 ,
\ea
\eeq
corresponding respectively to the condition that the coefficients of the positive and negative powers of $\cal E$ 
are zero in all orders. Equations (\ref{coeff+}) are identically satisfied using the definitions (\ref{def_pq}) and 
(\ref{def_w_tilde1}) of $p^{(N,j)},q^{(N,j)}$ and $\tilde w^{(j;11)}$. To show that also equations (\ref{coeff-}) 
are identically satisfied, we compare them with equations (\ref{g_chain}) and use again (\ref{def_pq}),  
(\ref{def_w_tilde1}), to finally derive the equation
\beq
\sum\limits_{k=1}^N \sum_{l=1}^k w^{(l)}_{x_k}\left(\sum\limits_{j=1}^{k-l}w^{(j)}\tilde w^{(k-l-j+1;1n)}-
\sum\limits_{j=0}^{k-l-1}\tilde w^{(k-l-j;11)}w^{(j+n)}\right)=0,~~n\ge 1
\eeq 
identically satisfied, due to (\ref{def_w_tilde1}) and (\ref{new}).

\subsection{Derivation of Calogero systems}
\label{Section:Calogero}
In the Lax pair of the S-integrable Calogero systems \cite{Calogero_syst1,Calogero_syst2,Calogero_syst3,Calogero_syst4,ZM_Calogero}, 
the spectral problem is 1-dimensional (like that of, say, KdV and  NLS), while the equation describing the evolution 
of the eigenfunction is a multidimensional PDE  in which an arbitrary  
number of additional independent variables are graded by powers of the spectral parameter; in addition, 
such spectral parameter satisfies a quasilinear PDE. 

It is therefore clear that Calogero systems combine properties  of the $S$-integrable PDEs in (1+1)-dimensions of \S \ref{Section:1+1} 
with properties of the multidimensional generalizations of the SDYM equation of \S 3.3 and, to generate them from equations 
(\ref{w_m}), we have to start with the matrix equation (\ref{H_sp}) and with evolutionary system (\ref{H_flows}) 
in which $\rho^{(mj1)}=\sigma^{(m2)}=0$:
\beq\label{eig}
w\Psi=\Psi\Lambda,
\eeq
\beq\label{H_flows_C}
\ba{l}
\Psi_{t_{m1}} - B^{(m1)} \Psi  \sigma^{(m1)}(\Lambda)=0,\;\;\; m\in\NN_+ ,\\
\Psi_{t_{m2}}+\sum_{j=1}^N \Psi_{x_{j2}}\rho^{(mj2)}(\Lambda) =0.
\ea
\eeq
Here we illustrate the construction of the simplest example of Calogero system, corresponding to  $N=m=1$ and 
$ \sigma^{(11)}(\Lambda)=\rho^{(112)}(\Lambda) =\Lambda$, $B=B^{(11)}$, using the notation $\tilde B=\tilde B^{(11)}$, $\tau=t_{11}$, 
$t=t_{12}$, $x=x_{12}$. Then the compatibility between the two equations (\ref{H_flows_C}) yields the following 
quasi-linear PDEs 
\begin{eqnarray}
\Lambda_{t}+ \Lambda_{x}\Lambda =0, \;\;\;\; \Lambda_{\tau}=0
\end{eqnarray}
for the matrix of eigenvalues; the compatibility between equations (\ref{eig}) and (\ref{H_flows_C}) yields instead the 
following matrix equations for field $w$
\begin{eqnarray}\label{wS1_C}
&&
w_{\tau}  + [w,B]w=0,\\\nonumber
&&
w_{t} + w_{x} w =0.
\end{eqnarray}
Using the Frobenius structure (\ref{W}) of $w$, we get the following discrete chains, for $n\in\NN_+$:
\begin{eqnarray}\label{C_s}
&&
w^{(n)}_{\tau} +[w^{(n+1)},\tilde B] +[w^{(1)},\tilde B] w^{(n)} =0,\\\nonumber
&&
w^{(n)}_{t} +w^{(1)}_{x}  w^{(n)} + 
 w^{(n+1)}_{x}  =0 ,
\end{eqnarray}
coinciding with eqs.(\ref{n_wave_ch_11}) for $k=1$,  and with (\ref{P_ch}a). Fixing $n=1$ in (\ref{C_s}), and 
eliminating $w^{(2)}$, one gets the following matrix PDE
\begin{eqnarray}\label{C_s2}
&&
[w^{(1)}_{t},\tilde B] + [ w^{(1)}_{x} w^{(1)},\tilde B] =w^{(1)}_{\tau x}  +\Big([w^{(1)},\tilde B] w^{(1)}\Big)_{x}.
\end{eqnarray}
Let $M=2$ and $\tilde B=\beta {\mbox{diag}}(1,-1)$; then eq.(\ref{C_s2}) yields
\begin{eqnarray}
&&
-\beta q_{t} -\frac{1}{2} q_{x \tau} -4\beta^2 q\partial^{-1}_{\tau}(qr)_{x}=0
,\\\nonumber
&&
\beta r_{t} -\frac{1}{2} r_{x \tau} -4\beta^2 r\partial^{-1}_{\tau}(qr)_{x}=0.
\end{eqnarray}
If  $r=\bar q$, $\beta=i$, then the above system becomes the following (2+1)-dimensional integrable variant of NLS:
\beq
iq_t+\frac{1}{2}q_{xx}-4q\partial^{-1}_{\tau}(q\bar q)_{x}=0,
\eeq
studied in \cite{Z_DS,Strachan}.

Using the dual of equations (\ref{sp_112_d}) with $m=k=1$, and equations (\ref{sp_ch_d}a,b),  
we derive the dual Lax pair ((\ref{sp2_t12_112}) for $k=1$ and (\ref{Sd_dual_Lax}a)) for (\ref{C_s2}):
\begin{eqnarray}
\label{sp2_t12_11_C}
&&
 \tilde\psi_{\tau}+{\cal{E}} \tilde\psi\tilde B+ \tilde\psi [\tilde B, w^{(1)}] =0,\\\nonumber
 &&
\tilde\psi_{t} + ({\cal{E}}\tilde\psi)_{x} = \tilde\psi w^{(1)}_{x}
\end{eqnarray}
and the corresponding Lax pair ((\ref{ch_sp2_t12_11}) for $k=1$ and (\ref{Lax_SD}a)):
\begin{eqnarray}
\label{ch_sp2_t12_11_C}
 &&
 \psi_{\tau}- \tilde B\psi{\cal{E}}-[\tilde B, w^{(1)}]\psi  =0
, \\\nonumber
&&
\psi_{t} + \psi_{x}{\cal{E}} + w^{(1)}_{x}\psi =0
\end{eqnarray}
for the system (\ref{C_s2}).

We end this section remarking that integrable PDEs associated with Lax pairs with varying spectral parameter have been studied 
also elsewhere, see, for instance, \cite{BZM}.

\subsection{Construction of solutions}
\label{Solutions_alg}
The construction of solutions for the three classes of $S$-integrable PDEs derived in \S 3.2, 3.3 and 
3.4 from eq.(\ref{w_m}), is based on the solution of the algebraic equations (\ref{Lambda},\ref{w_eigP_expl}), taking into account the  
block-matrix structure (\ref{Psi13}) of $\Psi$. Then the independent blocks $\Psi^{(ii)}$, $i=1,2,\dots$ are characterized by the 
explicit formula: 
 \begin{eqnarray}\label{Psi_gen}
\Psi^{(ii)}_{\alpha\beta}&=& F^{(ii)}_{\alpha\beta} \left(x_{11}-\sum_{m\ge 1}\rho^{(m11)}(\Lambda^{(i)}_\beta)t_{m1},\dots ,
x_{N1}-\sum_{m\ge 1}\rho^{(mN1)}(\Lambda^{(i)}_\beta)t_{m1};\right. \\\nonumber
&&\left.x_{12}-\sum_{m\ge 1}\rho^{(m12)}(\Lambda^{(i)}_\beta)t_{m2},\dots ,x_{N2}-\sum_{m\ge 1}\rho^{(mN2)}(\Lambda^{(i)}_\beta)t_{m2}
\right)\times
 \\\nonumber
&&
e^{\sum\limits_{k=1}^2\sum\limits_{m\ge 1} 
\tilde B^{(mk)}_{\alpha} \sigma^{(mk)}(\Lambda^{(i)}_\beta) t_{mk}},\;\;\alpha,\beta=1,\dots,M,\;\;i=1,2,\dots,
\end{eqnarray}
where $F^{(ii)}_{\alpha\beta}$ are arbitrary scalar functions of $2 N$ arguments (such that $F^{(ii)}_{\alpha\alpha}=1$), 
while the remaining blocks are given by the equations $\Psi^{(ij)}=\Psi^{(jj)}(\Lambda^{(j)})^{j-i}$. 
Once $\Lambda$ and $\Psi$ are constructed in this way,   
the blocks $w^{(i)}$ are obtained, from eq.(\ref{sol_Ch}), through the compact formula:
 \begin{eqnarray}\label{Sol_w}
 w^{(i)}_{\alpha\beta}=(\Psi \Lambda \Psi^{-1})_{\alpha(i M-M+\beta)},\;\;\alpha,\beta=1\dots,M.
 \end{eqnarray}

In the case of the (1+1)-dimensional $S$-integrable models of $\S$ \ref{Section:1+1}, corresponding to $\rho^{(ijk)}=0$, $\Lambda$ 
is an arbitrary constant diagonal matrix and formula (\ref{Psi_gen}) reduces to:
\beq\label{Psi_gen_1+1}
\Psi^{(ii)}_{\alpha\beta}=F^{(ii)}_{\alpha\beta}e^{\sum\limits_{k=1}^2\sum\limits_{m\ge 1} 
\tilde B^{(mk)}_{\alpha} \sigma^{(mk)}(\Lambda^{(i)}_\beta) t_{mk}},~~~~\alpha,\beta=1,\dots,M,\;\;i=1,2,\dots,
\eeq
where $F^{(ii)}_{\alpha\beta}$ are constant amplitudes. Then the solution (\ref{Sol_w}) is a rational combination of exponentials. 
It would be interesting to compare, in this case, the solution space generated by (\ref{Sol_w}) with that generated by 
Sato theory. 

In the case of the Calogero systems of $\S$ \ref{Section:Calogero}, when $\rho^{(mj1)}=\sigma^{(m2)} =0$, 
formula (\ref{Psi_gen}) reduces to:
\begin{eqnarray}\label{Psi_gen_C}
\Psi^{(ii)}_{\alpha\beta}&=& 
F^{(ii)}_{\alpha\beta} \left(x_{12}-\sum_{m\ge 1}\rho^{(m12)}(\Lambda^{(i)}_\beta)t_{m2},\dots ,x_{N2}-\sum_{m\ge 1}\rho^{(mN2)}(\Lambda^{(i)}_\beta)t_{m2}
\right)\times
 \\\nonumber
&&
e^{\sum\limits_{m\ge 1} 
\tilde B^{(m1)}_{\alpha} \sigma^{(m1)}(\Lambda^{(i)}_\beta) t_{m1}},\;\;\alpha,\beta=1,\dots,M,\;\;i=1,2,\dots.
\end{eqnarray}
where $F^{(ii)}_{\alpha\beta}$ are now arbitrary functions of $N$ arguments and $\Lambda$ is the implicit solution 
of the nondifferential equation 
   \begin{eqnarray}\label{Lambda_C}
\Lambda &=& E\left(x_{12}I-\sum_{m\ge1}\rho^{(m12)}(\Lambda)t_{m2},\dots ,x_{N2}I-\sum_{m\ge1}\rho^{(mN2)}(\Lambda)t_{m2}
\right),
\end{eqnarray}
following from the eq.(\ref{Lambda}), where $E$ is arbitrary diagonal matrix function of $N$ arguments.

\section{Generalizations}

The block Frobenius matrix (\ref{W}) is not the only possible structure  of $w$ allowing one to 
 generate  new types of integrable nonlinear PDEs, starting with $C$-integrable and $Ch$-integrable equations 
(\ref{C_hierarchy}) and (\ref{w_m}). A more general representation is given by the following block matrix:
\begin{eqnarray}\label{WW}
w=\left[\begin{array}{ccc}
W^{(11)}& W^{(12)}&\cdots\cr
W^{(21)}& W^{(22)}&\cdots\cr
\vdots&\vdots&\ddots
\end{array}\right],
\end{eqnarray} 
where each block $W^{(ij)}$ has one of the following two block matrix forms:
\begin{eqnarray}\label{WSC}
&&
W^{ij}=S^{(ij)}=\left[\begin{array}{cccc}
w^{(ij;1)}& w^{(ij;2)}& w^{(ij;3)}&\cdots\cr
I_M&0_M&0_M& \cdots\cr
0_M&I_M&0_M& \cdots\cr
\vdots&\vdots&\vdots&\ddots
\end{array}\right], \\\nonumber
&&
W^{ij}=C^{(ij)}=\left[\begin{array}{cccc}
w^{(ij;1)}& w^{(ij;2)}&w^{(ij;3)}&\cdots\cr
0_M& I_M&0_M& \cdots\cr
0_M&0_M&I_M& \cdots\cr
\vdots&\vdots&\vdots&\ddots
\end{array}\right],
\end{eqnarray}
and the blocks $w^{(ij;k)}$ are $M\times M$ matrices.

To provide consistency of this structure with the Lax pairs (\ref{Hopf},\ref{C_flows})
or (\ref{H_sp},\ref{H_flows}), we must take appropriate  structure
of matrices $B^{(nm)}$ and $\Psi$. 
Consider the simplest example of eqs.(\ref{Hopf},\ref{C_flows}) with $n=1$.
In this case $w$ satisfies the N-wave equations (\ref{n_wave}). Let
\begin{eqnarray}\label{WSC0}
w=\left[\begin{array}{ccc}
S^{(11)}& C^{(12)}\cr
C^{(21)}& C^{(22)}\cr
\end{array}\right].
\end{eqnarray}
Then $B^{(1m)}$  must be taken in the form
\begin{eqnarray}\label{B_S-C}
&&
B^{(1m)}={\mbox{diag}}(\hat B^{(m1)},\hat B^{(m2)}),\;\;
\hat B^{(mj)} = {\mbox{diag}}(\tilde  B^{(mj)},\tilde B^{(mj)}\dots),~~~~j=1,2,
\end{eqnarray}
where $\tilde  B^{(mj)}$ are diagonal matrices, and $\Psi$ must have the following block structure:
\begin{eqnarray}\label{Psi}
\Psi=\left[\begin{array}{ccc}
\Psi^{(11)}& \Psi^{(12)}\cr
\Psi^{(21)}& \Psi^{(22)}
\end{array}\right],
\end{eqnarray} 
where $\Psi^{(ij)}$  must satisfy the following system of linear PDEs (consequence of eq.(\ref{Hopf})), for $i\ge 2$:
\begin{eqnarray}\label{G_Psi1}
&&
\Psi^{(11;ij)}_x= \Psi^{(11;(i-1)j)} + \Psi^{(21;ij)},\\\label{G_Psi2}
&&
\Psi^{(12;ij)}_x= \Psi^{(12;(i-1)j)} + \Psi^{(22;ij)},\\\label{G_Psi3}
&&
\Psi^{(21;ij)}_x= \Psi^{(11;ij)} + \Psi^{(21;ij)},\\\label{G_Psi4}
&&
\Psi^{(22;ij)}_x= \Psi^{(12;ij)} + \Psi^{(22;ij)}.
\end{eqnarray}

In view of (\ref{WSC0}), eq.(\ref{n_wave})  reads ($t_i=t_{1i}$): 
 \begin{eqnarray}\label{S-C_11}
&&
S^{(11)}_{t_i} - \hat B^{(i1)} S^{(11)}_x + [S^{(11)}, \hat B^{(i1)}]S^{(11)}+
(C^{(12)}\hat B^{(i2)}-\hat B^{(i1)}C^{(12)}) C^{(21)} =0,\\\nonumber
&&
C^{(12)}_{t_i} - \hat B^{(i1)} C^{(12)}_x + [S^{(11)},\hat B^{(i1)}]C^{(12)}+
(C^{(12)} \hat B^{(i2)}-\hat B^{(i1)}C^{(12)}) C^{(22)} =0,\\\nonumber
&&
C^{(21)}_{t_i} - \hat B^{(i2)} C^{(21)}_x +
(C^{(21)} \hat B^{(i1)}-\hat B^{(i2)}C^{(21)}) S^{(11)}+
 [C^{(22)},\hat B^{(i2)}]C^{(21)} =0,\\\nonumber
&&
C^{(22)}_{t_i} - \hat B^{(i2)} C^{(22)}_x +
(C^{(21)} \hat B^{(i1)}-\hat B^{(i2)}C^{(21)}) C^{(12)} + [C^{(22)},\hat B^{(i2)}]C^{(22)}=0,
\end{eqnarray}
 where $i=1,2$.
Writing the block $(11)$ of each of these equations, one obtains
 \begin{eqnarray}\label{2_S-C_11}
&&
w^{(11;1)}_{t_i} - \tilde B^{(i1)} w^{(11;1)}_x + [w^{(11;1)},\tilde B^{(i1)}]w^{(11;1)}+ [w^{(11;2)},\tilde B^{(i1)}]+
(w^{(12;1)} \tilde B^{(i2)}-
\\\nonumber
&&
\tilde B^{(i1)}w^{(12;1)}) w^{(21;1)} =0,\\\nonumber
&&
w^{(12;1)}_{t_i} - \tilde B^{(i1)} w^{(12;1)}_x + 
[w^{(11;1)},\tilde B^{(i1)}]w^{(12;1)}+
(w^{(12;1)} \tilde B^{(i2)}-\tilde B^{(i1)}w^{(12;1)}) w^{(22;1)} =0,\\\nonumber
&&
w^{(21;1)}_{t_i} -  \tilde B^{(i2)} w^{(21;1)}_x +
(w^{(21;1)} \tilde B^{(i1)}-\tilde B^{(i2)}w^{(21;1)}) w^{(11;1)}+
(w^{(21;2)} \tilde B^{(i1)}-\tilde B^{(i2)}w^{(21;2)})
 +
 \\\nonumber
&&[w^{(22;1)},\tilde B^{(i2)}]w^{(21;1)} =0,\\\nonumber
&&
w^{(22;1)}_{t_i} -  \tilde B^{(i2)} w^{(22;1)}_x +
(w^{(21;1)} \tilde B^{(i1)}-\tilde B^{(i2)}w^{(21;1)}) w^{(12;1)} + [w^{(22;1)},\tilde B^{(i2)}]w^{(22;1)}=0.
\end{eqnarray}

Eliminating $w^{(11;2)}$ and $w^{(21;2)}$  from the eqs.(\ref{2_S-C_11}a,c) with $i=1,2$ and taking eqs.(\ref{2_S-C_11}b,d) with $i=2$, 
one obtains the  following (2+1)-dimensional evolutionary system of PDEs in the time variable $t_2$:
 \begin{eqnarray}\label{S-C_11_v}
&&
[{\cal{E}}^{(1)},\tilde B^{(21)}]-[{\cal{E}}^{(2)},\tilde B^{(11)}]=0,\\\label{S-C_11_q}
&&
q_{t_2} - \tilde B^{(21)} q_x + [v,\tilde B^{(21)}]q+(q \tilde B^{(22)}-\tilde B^{(21)}q) u =0,\\\label{S-C_11_E}
&&
E^{(1)}\tilde  B^{(21)} -\tilde B^{(22)}E^{(1)}-E^{(2)} \tilde B^{(11)} +\tilde B^{(12)}E^{(2)}=0,\\\label{S-C_11_u}
&&
u_{t_2} - \tilde B^{(22)} u_x +(w \tilde B^{(21)}-\tilde B^{(22)}w) q + [u,\tilde B^{(22)}]u=0,
\end{eqnarray}
where
\beq\label{def_E}
\ba{l}
{\cal{E}}^{(i)}=v_{t_i} -  \tilde B^{(i1)} v_x + [v,\tilde B^{(i1)}]v+(q \tilde B^{(i2)}-\tilde B^{(i1)}q) w,~~i=1,2, \\
E^{(i)}=w_{t_i}-\tilde B^{(i2)} w_x+(w \tilde B^{(i1)}-\tilde B^{(i2)}w) v+ [u,\tilde B^{(i2)}]w,~~i=1,2,
\ea
\eeq
and
\begin{eqnarray}
v=w^{(11;1)},\;\;q=w^{(12;1)},\;\;w=w^{(21;1)},\;\;u=w^{(22;1)},
\end{eqnarray}
supplemented by eqs.(\ref{2_S-C_11}b,d) with $i=1$: 
\begin{eqnarray}\label{constr1}
&&
q_{t_1} - \tilde B^{(11)} q_x + 
[v,\tilde B^{(11)}]q+
(q \tilde B^{(12)}-\tilde B^{(11)}q) u =0,\\\label{constr2}
&&
u_{t_1} -  \tilde B^{(12)} u_x +
(w \tilde B^{(11)}-\tilde B^{(12)}w) q + [u,\tilde B^{(12)}]u=0,
\end{eqnarray}
that can be considered as compatible constraints for the evolutionary system (\ref{S-C_11_v})-(\ref{S-C_11_u}).

The Lax pair, as well as the solution space for this system, can be obtained following
the procedures described in $\S$\ref{Section:S-C_sp} and $\S$\ref{Section:S-C_sol}.

If $w=q=u=0$, one obtains the $S$-integrable (2+1)-dimensional $N$-wave equation (\ref{S-C_11_v}) for $v$ (see 
eq.(\ref{S_n_w})). Instead, 
if  $v=w=q=0$, one obtains the $C$-integrable (1+1)-dimensional $N$-wave type equation (\ref{S-C_11_u}) or (\ref{constr2})
for $u$ (see eq.(\ref{n_wave})). Therefore equations (\ref{S-C_11_v})-(\ref{S-C_11_u}) and (\ref{constr1},\ref{constr2}) 
can be viewed as nonlinear PDEs sharing properties of S- and C- integrable systems. One can show that this property is shared   
by all the PDEs generated by reductions of the type (\ref{WW},\ref{WSC}); recovering, in particular, the ($n$+1)-dimensional 
($n>2$) nonlinear PDEs constructed in \cite{Z} by a generalization of the dressing method. 

\section{Summary and future perspectives}

We have established deep and remarkable connections among PDEs integrable by 
the inverse spectral transform method, the method of characteristics 
and the Hopf-Cole transformation. These relations can be used effectively to construct, for the generated 
S-integrable PDEs, the associated compatible systems of 
linear operators, their commuting flows and large classes of solutions. These results open several research perspectives. 
\begin{enumerate}
\item Use of the above derivation of the $S$ - integrable systems to investigate the corresponding space of analytic 
solutions generated from the seed solutions of the original C - and Ch - integrable PDEs. In particular, 
\begin{enumerate}
\item the connections between such solution space and that generated by Sato theory. In the KP case, the two solution 
spaces coincide; in other cases the connection is, at the moment, less clear.
\item The use of the quasi-linear PDEs for the eigenvalues to study in detail the wave breaking phenomena associated 
with  solutions of the SDYM equation and of its multidimensional generalizations.   
\end{enumerate}
\item Search for the integrable systems that should generate, through a suitable matrix reduction, the integrable 
PDEs equivalent to the commutation of vector fields, like equation (\ref{vec_nl}), a class of $S$-integrable systems 
not fitting yet into the general picture illustrated in this paper.
\item Generalization of the techniques presented in this paper to generate novel integrable systems, possibly 
in multidimensions. In particular, a systematic use of group theory tools to explore reductions different from 
the block Frobenius matrix one.  
\item Construction of the discrete analogue of the results of this paper. In this respect, we remark that, while 
the discretization of the 
results of \S 2 does not present, in principle, any conceptual problem, and will be the subject of a subsequent paper, 
the discretization of the results of \S 3, if $\rho^{(ijk)}\ne 0$, is not clear, since a satisfactory 
discretization of the method of characteristics and of equation (\ref{w}) for $\rho^{(i)}\ne 0$, in the scalar and matrix cases, 
are, at the moment, unknown.  

\end{enumerate}

\vskip 10pt \noindent {\bf Acknowledgments}. 
AIZ was supported by the RFBR grants 07-01-00446, 06-01-92053, 06-01-90840, by
the grant NS 7550.2006.2 and by the INFN grant 2007. AIZ thanks Prof. A.B.Shabat for useful discussion.


\end{document}